\documentclass[useAMS,usenatbib]{mn2e}

\usepackage{mathptmx}
\usepackage{amssymb}
\usepackage[fleqn]{amsmath}
\usepackage{url}
\usepackage{graphicx}
\usepackage{enumitem}
\usepackage{txfonts}
\usepackage{bibentry} 
\usepackage{etoolbox} 

\usepackage{xspace}

\newcommand{\SimpleX}{\textsc{simplex}\xspace}
\newcommand{\MdotA}{\ifmmode{\dot{M}_{\eta_{\mathrm{A}}}}\else{$\dot{M}_{\eta_{\mathrm{A}}}$}\fi\xspace}
\newcommand{\MdotB}{\ifmmode{\dot{M}_{\eta_{\mathrm{B}}}}\else{$\dot{M}_{\eta_{\mathrm{B}}}$}\fi\xspace}
\newcommand{\etaA}{$\eta_{\mathrm{A}}$\xspace}
\newcommand{\etaB}{$\eta_{\mathrm{B}}$\xspace}
\newcommand{\ec}{$\eta$~Car\xspace}
\makeatletter
\newcommand\bibstyle@comma{\bibpunct{(}{)}{,}{a}{}{,}}
\newcommand\bibstyle@semicolon{\bibpunct{(}{)}{;}{a}{}{,}}
\makeatother
\pretocmd\citet{\citestyle{comma}}\relax\relax
\pretocmd\citep{\citestyle{semicolon}}\relax\relax

\newcommand{\Ms}{\ifmmode{~\mathrm{M}_{\odot}}\else{$\mathrm{M}_{\odot}$}\fi\xspace}
\newcommand{\Msy}{\ifmmode{\Ms \per{yr}}\else {$\Ms \per{yr}$}\fi\xspace}
\newcommand{\Ls}{\ifmmode{~\mathrm{L}_{\odot}}\else{$\mathrm{L}_{\odot}$}\fi\xspace}
\newcommand{\kms}{\ifmmode{~\mathrm{km}\per{s}}\else {$\mathrm{km}\per{s}$}\fi\xspace}
\newcommand{\Rs}{\ifmmode{~\mathrm{R}_{\odot}}\else{$\mathrm{R}_{\odot}$}\fi\xspace}
\newcommand{\per}[1]{\ifmmode{\mathrm{\,#1}^{-1}}\else {$\mathrm{\,#1}^{-1}$}\fi\xspace}

\newcommand{\ten}[1]{\ifmmode{10^{#1}}\else{$10^{#1}$}\fi\xspace}
\newcommand{\sci}[2]{\ifmmode{#1 \times 10^{#2}}\else{$#1 \times 10^{#2}$}\fi\xspace}

\newcommand{\HI}{\ifmmode{\mathrm{H\,I}}\else{H\textsc{$\,$i}}\fi\xspace}
\newcommand{\HII}{\ifmmode{\mathrm{H\,II}}\else{H\textsc{$\,$ii}}\fi\xspace}
\newcommand{\HeI}{\ifmmode{\mathrm{He\,I}}\else{He\textsc{$\,$i}}\fi\xspace}
\newcommand{\HeII}{\ifmmode{\mathrm{He\,II}}\else{He\textsc{$\,$ii}}\fi\xspace}
\newcommand{\HeIII}{\ifmmode{\mathrm{He\,III}}\else{He\textsc{$\,$iii}}\fi\xspace}
\newcommand{\nuHI}{\ifmmode{\nu_{\ion{H}{0+}}}\else{$\nu_{\ion{H}{0+}}$}\fi\xspace}
\newcommand{\nuHeI}{\ifmmode{\nu_{\ion{He}{0+}}}\else{$\nu_{\ion{He}{0+}}$}\fi\xspace}
\newcommand{\nuHeII}{\ifmmode{\nu_{\ion{He}{+}}}\else{$\nu_{\ion{He}{+}}$}\fi\xspace}
\newcommand{\ion}[2]{\ifmmode{\mathrm{#1}^{#2}}\else{#1$^{#2}$}\fi\xspace}

\defcitealias{Madura_etA_2013}{M13}
\defcitealias{Clementel_etA_2014}{C14}

\usepackage{color}

\usepackage{ifdraft}

\ifdraft{\newcommand{\rmf}{\comment}} {\newcommand{\rmf}{}}

\title[\ec He ionization structure at apastron]{3D radiative transfer simulations of Eta Carinae's inner colliding winds -- I. Ionization structure of helium at apastron}

\author[Clementel et al.]{N. Clementel$^{1}$\thanks{E-mail: clementel@strw.leidenuniv.nl}, T.~I. Madura$^{2}$, C.~J.~H. Kruip$^{1}$, J.-P. Paardekooper$^{3,4}$ and T.~R. Gull$^{2}$\\
$^{1}$Leiden Observatory, Leiden University, PO Box 9513, 2300 RA Leiden, the Netherlands\\
$^{2}$Astrophysics Science Division, Code 667, NASA Goddard Space Flight Center, Greenbelt, MD 20771, USA\\
$^{3}$Zentrum f\"ur Astronomie, Institut f\"ur Theoretische Astrophysik, Universit\"at Heidelberg, Alber--Ueberle--Str. 2, D-69120 Heidelberg, Germany\\
$^{4}$Max Planck Institute for Extraterrestrial Physics, PO Box 1312, Giessenbachstr., D--85741 Garching, Germany\\
}

\begin{document}

\date{Accepted 2014 December 8. Received 2014 December 8; in original form 2014 October 2}

\pagerange{\pageref{firstpage}--\pageref{lastpage}} \pubyear{2014}

\maketitle

\label{firstpage}

\begin{abstract}
The highly eccentric binary system Eta Carinae (\ec) shows numerous time-variable emission and absorption features. These observational signatures are the result of interactions between the complex three-dimensional (3D) wind--wind collision regions and photoionization by the luminous stars. Specifically, helium presents several interesting spectral features that provide important clues on the geometry and physical properties of the system and the individual stars. We use the \SimpleX algorithm to post-process 3D smoothed particle hydrodynamics simulation output of the interacting winds in \ec in order to obtain the fractions of ionized helium assuming three different primary star (\etaA) mass-loss rates. The resultant ionization maps constrain the regions where helium is singly- and doubly-ionized. We find that reducing \etaA's mass-loss rate (\MdotA) increases the volume of \ion{He}{+}. Lowering \MdotA produces large variations in the volume of \ion{He}{+} in the pre-shock \etaA wind on the periastron side of the system. Our results show that binary orientations in which apastron is on our side of the system are more consistent with available observations. We suggest that small variations in \MdotA might explain the observed increase in \HeI absorption in recent decades, although numerous questions regarding this scenario remain open. We also propose that the absence of broad \HeI lines in the spectra of \ec between its 1890's eruption and $\sim$1944 might be explained by \etaB's \ion{He}{0+}-ionizing photons not being able to penetrate the wind-wind interaction region, due to a higher \MdotA at that time (by a factor $\gtrsim$2, compared to the present value).
\end{abstract}

\begin{keywords}
hydrodynamics -- radiative transfer -- binaries: close -- stars: individual: Eta Carinae -- stars: mass-loss -- stars: winds, outflows
\end{keywords}


\section{Introduction}\label{intro}

Eta Carinae (\ec) is probably most famous for its `Great Eruption' in the 1840s, when it temporarily became the second brightest non-solar-system object in the sky and ejected $\sim$10--40~\Ms, forming the dusty bipolar `Homunculus' nebula \citep{Davidson_Humphreys_1997, Smith_etA_2003a, Gomez_etA_2010, Steffen_etA_2014}. Near the centre of the Homunculus lies \ec itself, an extremely luminous ($L_{\mathrm{Total}} \gtrsim \sci{5}{6} \Ls$) and highly eccentric ($e \sim 0.9$) binary with a 5.54~yr orbit \citep{Damineli_etA_1997, Hillier_etA_2001, Damineli_etA_2008b, Damineli_etA_2008a, Corcoran_etA_2010}. The primary component, \etaA, is a Luminous Blue Variable (LBV) and our closest example of a supermassive star ($D = 2.3$~kpc, $M_{\star}\sim 100 \Ms$, $T_{\mathrm{eff}} \simeq 9400$~K; \citealt{Hillier_etA_2001,Smith_2006}). The secondary, \etaB, is thought to be a hotter ($T_{\mathrm{eff}} \simeq$~36,000--41,000~K), but less luminous ($L_{\star}/ \Ls \approx$~\ten{5}--\ten{6}), O- or Wolf Rayet-type star \citep{Pittard_Corcoran_2002, Verner_etA_2005, Hillier_etA_2006, Teodoro_etA_2008, Mehner_etA_2010}.

Because they are so luminous, both components of \ec have powerful radiation-driven stellar winds. Multiwavelength observations obtained over the last two decades \citep{Corcoran_2005, Hamaguchi_etA_2007, Damineli_etA_2008a, Henley_etA_2008, Groh_etA_2010b, Corcoran_etA_2010, Gull_etA_2009, Gull_etA_2011, Teodoro_etA_2013} indicate that \etaA's slow, extremely dense wind ($v_{\infty} \approx 420 \kms$, $\MdotA \approx \sci{8.5}{-4} \Msy$; \citealt{Hillier_etA_2001, Groh_etA_2012a}) collides with \etaB's less dense ($\MdotB \approx \sci{1.4}{-5} \Msy$), but much faster ($v_{\infty} \approx 3000 \kms$; \citealt{Pittard_Corcoran_2002, Parkin_etA_2009}), wind. This wind--wind collision (WWC) produces the shock-heated gas responsible for the observed time-variable 2--10~keV X-ray emission \citep{Pittard_Corcoran_2002, Corcoran_2005, Hamaguchi_etA_2007, Okazaki_etA_2008, Corcoran_etA_2010, Parkin_etA_2009, Parkin_etA_2011, Hamaguchi_etA_2014} that is a key signature of a colliding wind binary \citep{Luo_etA_1990, Stevens_etA_1992}.

The WWC, orbital motion, and presence of \etaB lead to numerous other forms of time-variable emission and absorption seen across a wide range of wavelengths \citep[see e.g.][]{Damineli_etA_2008b}. Observational signatures that arise as a result of the WWC and \etaB's ionizing radiation are important for studying \ec as they provide crucial information about the physical properties of the stars and the system as a whole. Three-dimensional (3D) hydrodynamical simulations show that the fast wind of \etaB has a significant impact on shaping the wind of \etaA \citep{Okazaki_etA_2008, Madura_2010, Parkin_etA_2011, Madura_Groh_2012, Madura_etA_2012, Madura_etA_2013, Russell_2013}, affecting greatly the observed optical and ultraviolet (UV) spectra of the system, as well as the interpretation of various line profiles and interferometric observables \citep{Groh_etA_2010a, Groh_etA_2010b, Groh_etA_2012a, Groh_etA_2012b}. Recently, \citet[][hereafter C14]{Clementel_etA_2014} presented 3D radiative transfer (RT) simulations that illustrate the effects of \etaB's ionizing radiation on the outer regions of \ec's extended ($r \approx 1500$~au) colliding winds. However, to date there has been no detailed 3D RT modelling to determine the effects of \etaB's ionizing radiation on \etaA's inner wind, the inner wind--wind interaction region (WWIR), or the numerous observed emission and absorption lines that arise in the inner $\sim$150~au of the system.

A very important series of spectral features that have fascinated and perplexed researchers of \ec for decades are those due to helium. It was the periodic variation of the \HeI $\lambda$10830 emission line that originally led to the discovery of binarity in \ec \citep{Damineli_1996, Damineli_etA_1997}, and it is the disappearance of the narrow emission-line component of \HeI $\lambda$6678 that is typically used to define the starting point of a 5.54~yr spectroscopic cycle \citep{Damineli_etA_2008a}. Present-day broad wind lines of \HeI, most notably $\lambda$7067, are thought to be excited by the UV radiation of \etaB and arise somewhere in/near the WWIR between the stars \citep{Nielsen_etA_2007, Damineli_etA_2008a}. Since the \HeI lines are recombination lines, they are produced in regions of \ion{He}{+} rather than regions of neutral He. The locations of the strongest broad \HeI emission features are spatially unresolved in \emph{Hubble Space Telescope} data, and almost certainly originate less than $\sim$100--200~au from \etaA \citep{Humphreys_etA_2008}. The broad \HeI emission lines are consistently blueshifted throughout most of the 5.54~yr orbit and exhibit an interesting double peak profile that varies in intensity and velocity, especially across periastron passage \citep{Nielsen_etA_2007}.

The broad \HeI P~Cygni absorption components also vary in velocity and strength over \ec's entire 5.54~yr period. The \HeI absorption is strongest in the two-year interval centred on periastron, and relatively weak at other phases \citep{Nielsen_etA_2007}. The absorption is always blueshifted, and hence must be produced by material between the observer and the continuum source \etaA, although there is still some debate over whether the absorption is directly related to material in the WWIR \citep{Damineli_etA_2008a} or produced by the pre-shock wind of \etaA \citep{Nielsen_etA_2007}.

Further complicating the story is the observed gradual increase in the amount of P~Cygni absorption over the last $\sim$10~yr. Since 1998, and most especially after the 2009 event, the strength of the \HeI absorption has increased compared to similar phases of previous cycles, while the emission strength has remained essentially unchanged \citep{Groh_Daminelli_2004, Mehner_etA_2010, Mehner_etA_2012}. \citet{Mehner_etA_2012} attribute this and other recent observed changes to a gradual decrease of \etaA's mass-loss rate by a factor of $\sim$2--3 between 1999 and 2010. It is hypothesized that this decrease in \etaA mass-loss rate led to important changes in the ionization structure of \etaA's wind and the WWIRs, as caused by the presence of \etaB (see e.g. fig.~5 of \citealt{Mehner_etA_2012}). However, this idea has yet to be quantitatively tested or modelled, and the results of 3D smoothed particle hydrodynamic (SPH) simulations appear to argue against such a large, gradual change in \etaA's mass-loss rate \citep[see][hereafter M13]{Madura_etA_2013}.

Finally, there is the mystery regarding the origin of \ec's \HeII $\lambda$4686 emission, which is only seen strongly around periastron (between phases $\sim 0.98$ and 1.03; \citealt{Steiner_Daminelli_2004, Martin_etA_2006, Mehner_etA_2011, Teodoro_etA_2012}). The locations and physical mechanisms that give rise to this emission in \ec are not completely agreed upon, although several possible scenarios have been proposed (\citealt{Steiner_Daminelli_2004, Martin_etA_2006, Mehner_etA_2011, Teodoro_etA_2012}; \citetalias{Madura_etA_2013}). The key difficulty with determining which scenario, if any, is correct is the lack of any detailed quantitative modelling of the 3D ionization structure of the innermost stellar winds and WWIR.

The goal of this paper is to compute full 3D RT simulations of the effects of \etaB's ionizing radiation on \ec's inner winds and WWIR, focusing on the ionization structure of helium at orbital phases around apastron (i.e. the spectroscopic high state; \citealt{Gull_etA_2009}). The ionization structure of helium at periastron (the spectroscopic low state) is investigated in a subsequent paper (Clementel et al. MNRAS, submitted). We apply the \SimpleX algorithm for 3D RT on an unstructured Delaunay grid \citep{Ritzerveld_Icke_2006, Ritzerveld_2007, Kruip_etA_2010, Paardekooper_etA_2010, Paardekooper_etA_2011} to recent 3D SPH simulations of \ec's binary colliding winds that include orbital motion, radiative cooling, and radiative forces \citepalias{Madura_etA_2013}. Using \SimpleX, we obtain detailed ionization fractions of helium at the resolution of the original SPH simulations. This should help us determine much more precisely where, and to what extent, the various emission and absorption components of the observed broad helium lines can form. This paper lays the foundation for future work aimed at generating synthetic spectra for comparison to observational data. We note that we focus solely on interpreting the \emph{broad} emission and absorption features of helium that arise in the stellar winds and WWIRs, and not the much narrower ($\lesssim 50$~\kms) features that form in the Weigelt blobs and other dense, slow-moving near-equatorial circumstellar ejecta \citep{Weigelt_Ebersberger_1986, Damineli_etA_2008a}.

We describe our numerical approach, including the SPH simulations, the \SimpleX code, and the RT simulations in Section~\ref{sec:Method}. Section~\ref{sec:Results} describes the results. A discussion of the results and their implications is in Section~\ref{sec:Discussion}. Section~\ref{sec:Summary} summarizes our conclusions and outlines the direction of future work.


\section{Methods}\label{sec:Method}

\subsection{The 3D SPH simulations}\label{ssec:SPH}

The hydrodynamical simulations used in this work correspond to the three small-domain ($r = 10\,a = 155$~au) 3D SPH simulations of \citetalias{Madura_etA_2013}. This computational domain size was chosen in order to investigate, at sufficiently high resolution, the structure of \ec's inner WWIRs and their effects on \etaB's ionizing radiation since the `current' interaction between the two winds occurs at spatial scales comparable to the semi-major axis length $a \approx 15.4$~au $\approx 0.0067$~arcsec at $D = 2.3$~kpc. In the following, we briefly describe only the essential aspects of the SPH code and setup. We refer the reader to \citetalias{Madura_etA_2013} and references therein for further details.

Radiative cooling is implemented using the Exact Integration Scheme of \citet{Townsend_2009}, with the radiative cooling function $\Lambda(T)$ calculated using \textsc{cloudy}~90.01 \citep{Ferland_etA_1998} for an optically thin plasma with solar abundances. The pre-shock stellar winds and rapidly-cooling dense gas in the WWIRs are assumed to be maintained at a floor temperature $= \ten{4}$~K due to photoionization heating by the stars \citep{Parkin_etA_2011}. Radiative forces are incorporated via the `antigravity' formalism described in \citetalias{Madura_etA_2013} and \citet{Russell_2013}. We parametrize the stellar winds using the standard `beta-velocity law' $v(r) = v_{\infty}(1-R_{\star}/r)^{\beta}$, where $v_{\infty}$ is the wind terminal velocity, $R_{\star}$ the stellar radius, and $\beta$ ($= 1$) a free parameter describing the steepness of the velocity law. Effects due to radiative braking \citep{Gayley_etA_1997, Parkin_etA_2011}, photospheric reflection \citep{Owocki_2007}, and self-regulated shocks \citep{Parkin_Sim_2013}, are not included since such effects are not expected to play a prominent role in \ec (\citealt{Parkin_etA_2009, Parkin_etA_2011, Russell_2013}; \citetalias{Madura_etA_2013}). We include the more important velocity-altering effects of `radiative inhibition' \citep{Stevens_Pollock_1994, Parkin_etA_2009, Parkin_etA_2011}. However, possible changes to the mass-loss rates due to radiative inhibition are not included. These are not expected to be significant in \ec and should not greatly affect our results or conclusions \citepalias{Madura_etA_2013}.

We use a standard $xyz$ Cartesian coordinate system and set the orbit in the $xy$ plane, with the origin at the system centre of mass and the major axis along the $x$-axis. The stars orbit counter-clockwise when viewed from along the $+z$-axis. By convention, $t = 0$ ($\phi = t/2024 = 0$) is defined as periastron. Simulations are started at apastron and run for multiple consecutive orbits.

The outer spherical simulation boundary is set at $r = 10\,a$ from the origin. Particles crossing this boundary are removed from the simulations. The adopted simulation parameters (Table~\ref{tab:Tab1}) are consistent with those derived from the available observations, although there is some debate on the present-day value of \MdotA (see \citetalias{Madura_etA_2013} for details). In an attempt to better constrain \MdotA, \citetalias{Madura_etA_2013} performed a series of 3D SPH simulations assuming three different \MdotA. We use the same naming convention as \citetalias{Madura_etA_2013} when referring to the SPH and \SimpleX simulations in this paper, namely, Case~A ($\MdotA = \sci{8.5}{-4}$~\Msy), Case~B ($\MdotA = \sci{4.8}{-4}$~\Msy), and Case~C ($\MdotA = \sci{2.4}{-4}$~\Msy).

\begin{table}
\caption{Stellar, wind, and orbital parameters of the 3D SPH simulations}
\label{tab:Tab1}
\begin{center}
\begin{tabular}{l c c}\hline
  Parameter & \etaA & \etaB \\ \hline
  $M_{\star}$ (\Ms) & 90 & 30 \\
  $R_{\star}$ (\Rs) & 60 & 30 \\
  $\dot{M}$ (\ten{-4} \Msy) & 8.5, 4.8, 2.4 & 0.14 \\
  $v_{\infty}$ (\kms) & 420 & 3000 \\
  $\eta$ & \multicolumn{2}{c}{0.12, 0.21, 0.42} \\
  $P_{\mathrm{orb}}$ (d) & \multicolumn{2}{c}{2024} \\
  $e$ & \multicolumn{2}{c}{0.9} \\
  $a$ (au) & \multicolumn{2}{c}{15.45} \\ \hline
\end{tabular}
\end{center}
\textbf{Notes:} $M_{\star}$ and $R_{\star}$ are the stellar mass and radius. $\dot{M}$ and $v_{\infty}$ are the stellar-wind mass-loss rate and terminal speed, respectively. $\eta \equiv (\dot{M} v_{\infty})_{\eta_{\mathrm{B}}}/(\dot{M} v_{\infty})_{\eta_{\mathrm{A}}}$ is the secondary/primary wind momentum ratio, $P_{\mathrm{orb}}$ is the orbital period, $e$ is the eccentricity, and $a$ is the length of the orbital semimajor axis.
\end{table}


\subsection{The \SimpleX algorithm for RT on an unstructured mesh}\label{ssec:SimpleX}

For the RT calculations, we post-process the 3D SPH simulation output using the \SimpleX algorithm \citep{Ritzerveld_Icke_2006, Ritzerveld_2007, Kruip_etA_2010, Paardekooper_etA_2010, Kruip_2011}. We employ a methodology nearly identical to that in \citetalias{Clementel_etA_2014}, but use an updated version of \SimpleX, which contains several important improvements over the version used in \citetalias{Clementel_etA_2014}. We discuss the relevant differences in the following sections, and briefly describe the key aspects of the code and their relevance to this work. We refer the reader to \citetalias{Clementel_etA_2014} and references therein for further details on \SimpleX and its applications.

\subsubsection{Grid construction and density distribution}\label{ssec:Grid}

\begin{figure*}
 \begin{center}
   \rmf{\includegraphics[width=174mm]{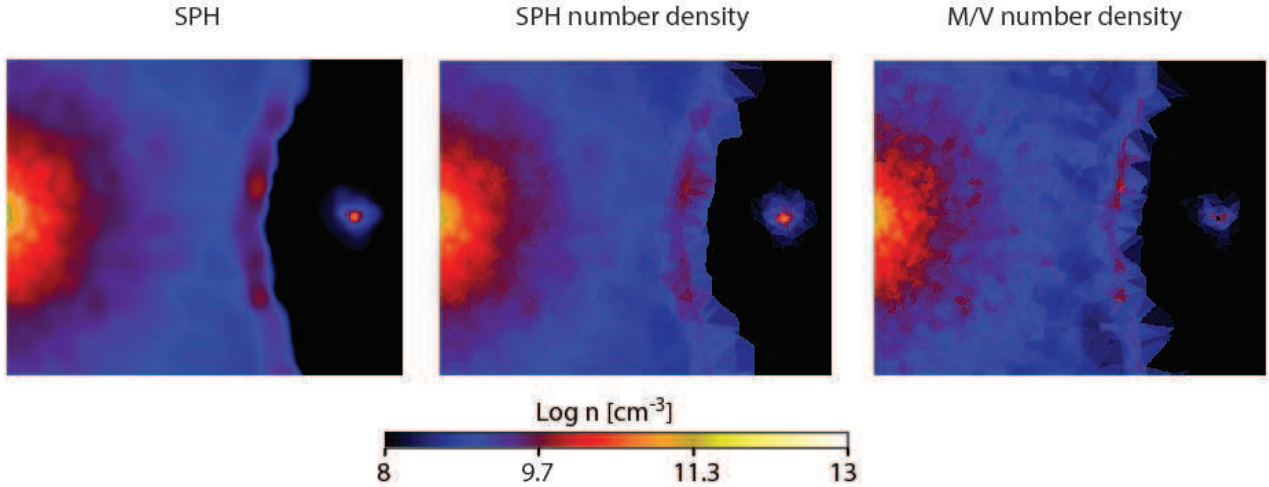}}
   \caption{Zoom of the central region for a slice in the $xy$ orbital plane through the 3D simulation volume for the Case~A simulation at apastron. Colour shows number density on a logarithmic scale (cgs units) using three different visualization approaches (see Section~\ref{ssec:Grid}). Left-hand panel: direct visualization using \textsc{splash}. Middle panel: SPH kernel approach. Right-hand panel: SPH particle mass divided by Voronoi cell volume approach.}\label{fig:MoverV_vs_SPH}
 \end{center}
\end{figure*}

\begin{figure*}
 \begin{center}
    \rmf{\includegraphics[width=174mm]{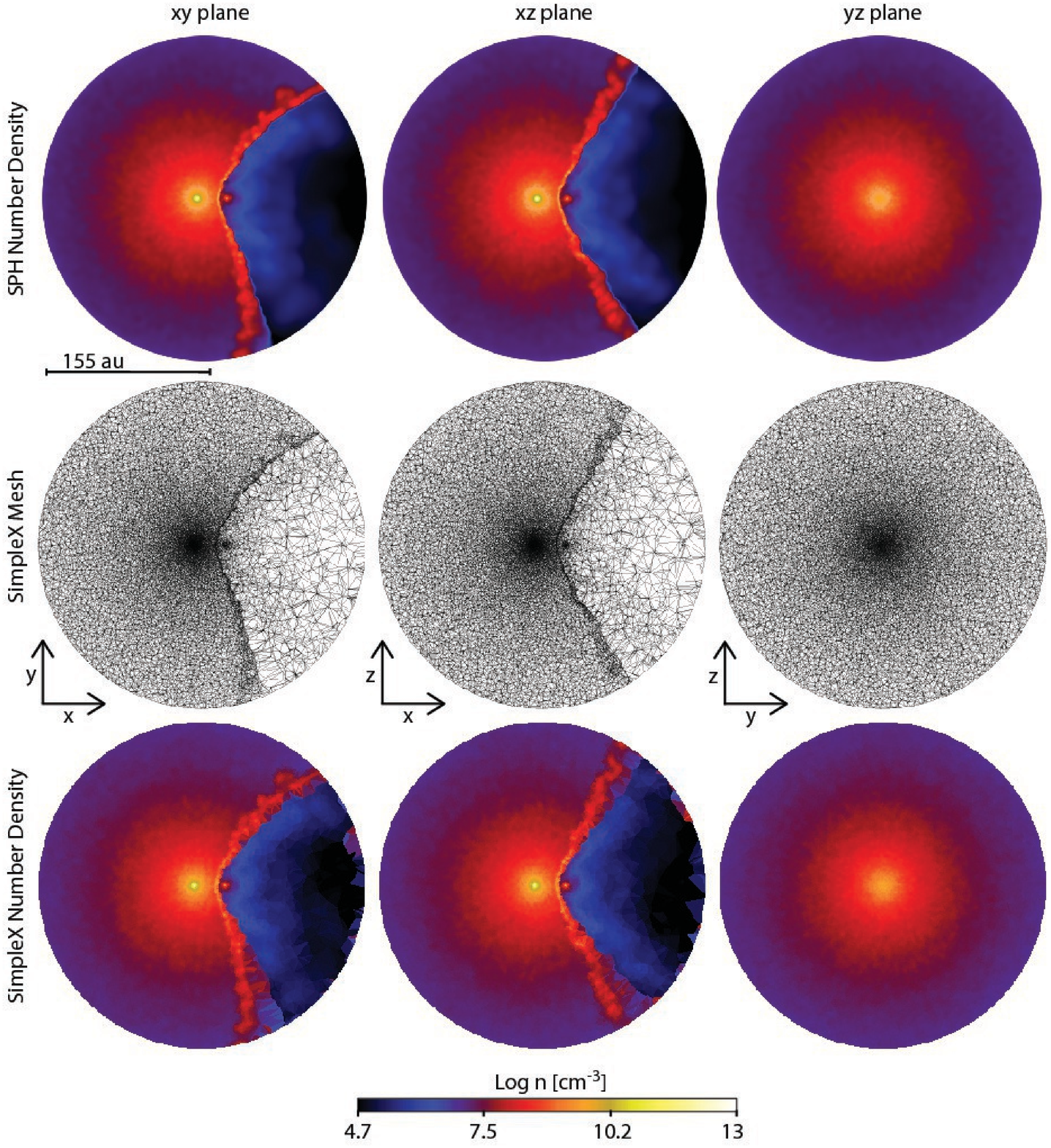}}
    \caption{Slices in the $xy$ (left-hand column), $xz$ (middle column) and $yz$ (right-hand column) planes through the 3D simulation volume for the Case~A simulation at apastron. Rows show, from top to bottom, the original SPH number density distribution (log scale, cgs units), the \SimpleX mesh, and the resulting \SimpleX number density (same log scale, cgs units). The resolution of the \SimpleX mesh, as well the number density, follow well the resolution of the original SPH data. The length-scale is shown under the top-left panel. In the first column (i.e. the orbital plane) \etaA is to the left and \etaB is to the right.}\label{fig:dens_mesh}
 \end{center}
\end{figure*}

As in \citetalias{Clementel_etA_2014}, we use the SPH particles themselves as the generating nuclei for the Voronoi--Delaunay mesh. We assign to the nucleus of each Voronoi cell the corresponding SPH density, computed using the standard SPH cubic spline kernel \citep{Monaghan_1992}. This helps ensure that the number density used in the \SimpleX calculations closely matches that of the original SPH simulations. Note that this approach differs from that used by \citetalias{Clementel_etA_2014}, wherein the \SimpleX density is obtained by dividing the SPH particle mass by the corresponding Voronoi cell volume. Using the SPH kernel produces smoother densities than the Voronoi cell-volume approach, since the SPH kernel samples a larger number of particles over a larger volume, resulting in densities that are less affected by local differences in the SPH particle distribution (see Fig.~\ref{fig:MoverV_vs_SPH}). Comparison with a direct visualization of the SPH density output (using \textsc{splash}, \citealt{Price_2007}, left-hand panel of Fig.~\ref{fig:MoverV_vs_SPH}) shows that the SPH kernel approach indeed matches better the density distribution of the original SPH simulations. Fig.~\ref{fig:dens_mesh} shows an example of the \SimpleX mesh and number density at apastron for a representative 3D SPH simulation of \ec.

\subsubsection{Ionization state and chemistry of the gas}\label{ssec:Chemistry}

As in \citetalias{Clementel_etA_2014}, we perform the RT calculations in post-processing. We consider the ionization of hydrogen and helium atoms by both photo- and collisional ionization. The ionization rate equations are solved on a time-step smaller than the RT time-step to ensure photon conservation \citep{Pawlik_Schaye_2008, Paardekooper_etA_2010}. This can lead to very small time-steps in cells where the photoionization time-scale is very small. To speed up the computation in these cells, we instead use the time-averaged optical depth to compute the photoionization rate and iterate for convergence \citep{Mellema_etA_2006, Friedrich_etA_2012}.

\subsubsection{Treatment of the ionizing spectrum and transport method}\label{ssec:SimpleX3}

In numerical simulations involving radiation it is necessary to approximate the continuous spectrum of radiation with a finite number of discrete frequency bins due to memory requirements. In \citetalias{Clementel_etA_2014}, the extreme limit of a single frequency bin, commonly referred to as the `grey approximation', was used. Although in the grey approximation all spectral information is lost, it is still possible to enforce the conservation of a quantity of importance such as the number of ionizations per unit time or the energy deposition into the medium per unit time. However, since in this work we are interested in the detailed ionization structure of He, and we want to capture the behaviour of the product of the spectrum and cross-sections in sufficient detail, we now employ three frequency bins.

The width of each frequency bin is set by the ionization energy of each species. The first bin ranges from the ionization frequency of \ion{H}{0+} (\nuHI = \sci{3.28}{15}~Hz) to that of \ion{He}{0+} (\nuHeI = \sci{5.93}{15}~Hz), the second from \nuHeI to \nuHeII (\sci{1.31}{16}~Hz), and the third from \nuHeII to a maximum frequency equal to 10 times \nuHI. We use an effective cross-section representation to determine the correct number of absorptions within each frequency bin. In this case, the limits of integration in equations~7 and 8 of \citetalias{Clementel_etA_2014} are over the frequency range of the bin of interest.

In \citetalias{Clementel_etA_2014}, photons were transported across the \SimpleX grid using ballistic transport. With this method, the incoming direction of the photons is used to define the outgoing direction, and the outgoing photons are distributed in 3D over the three most forward edges of the Delaunay triangulation. One drawback of this approach is that, due to the random nature of the outgoing directions in the Delaunay grid, the radiation may lose track of the original incoming direction after many steps. If the cells are optically thin, this can result in a radiation field that is too diffusive, leading to overestimates of the ionization fractions. To solve this problem, the original direction of the photons is preserved by confining them to solid angles corresponding to global directions in space. This is known as direction-conserving transport \citep[DCT;][]{Kruip_etA_2010, Paardekooper_2010}.

In DCT, if a photon is emitted in a certain direction associated with a solid angle, it will remember this direction and, unless it interacts with atoms on the grid, it will stay in the same solid angle as it travels through the grid. This effectively decouples the directionality of the radiation field from the directions present in the grid. In this work, we use 84 directions with DCT, implying a solid angle of $\mathrm{\pi}$/21~sr for each unit vector.

\subsubsection{The ionizing source \etaB}\label{ssec:Source}

Based on \citet{Mehner_etA_2010}, \citet{Verner_etA_2005} and \citet{Madura_etA_2012}, we consider \etaB to be an O5 giant with $\mathrm{T_{eff}}\approx 40,000$~K. We assume a total ionizing flux for H and He of \sci{3.58}{49} photons\per{s} \citep{Martins_etA_2005}.

As described in Section~\ref{ssec:SimpleX3}, we use three bins to sample the spectrum, which we approximate with a blackbody. We are therefore interested in the number of ionizing photons in each bin, $\mathrm{photons}_{i}$, which depends on the blackbody temperature:

\begin{equation}\label{eq:photons_i}
\mathrm{photons}_{i}= \int^{\lambda_{i}}_{0} B_{\lambda}(T) \mathrm{d}\lambda,
\end{equation}\\
where the $\lambda_{i}$ are the limiting wavelengths for the ionization of \ion{H}{0+} ($\lambda = 912$~\AA), \ion{He}{0+} ($\lambda = 504$~\AA) and \ion{He}{+} ($\lambda = 228$~\AA), and $B_{\lambda}$ is the Planck spectrum

\begin{equation}\label{eq:planck}
B_{\lambda}(T) \propto (\lambda^5  \exp[hc/(\lambda k_{B} T)])^{-1}.
\end{equation}

In \citet{Martins_etA_2005}, the fluxes able to ionize H and He are defined as

\begin{equation}\label{eq:q_i}
q_{i} = \int^{\lambda_{i}}_{0} \frac{\mathrm{\pi} \lambda F_{\lambda}}{hc} \mathrm{d}\lambda,
\end{equation}\\
where $F_{\lambda}$ is the flux expressed in $\mathrm{erg/s/cm^{2}/}$\AA. We set the blackbody temperature to the value that produces the correct ratio $\mathrm{photons}_{\ion{H}{0+}} / \mathrm{photons}_{\ion{He}{0+}}$ (in this case $T_{\mathrm{bb}} = 49,000$~K), in accordance to the $q_{\ion{H}{0+}}/q_{\ion{He}{0+}}$ ratio in \citet{Martins_etA_2005}. $q_{\ion{He}{+}}$ is effectively zero for \etaB.

\subsubsection{Visualization of the unstructured mesh results}\label{ssec:VisIt}

\begin{figure*}
 \begin{center}
   \rmf{\includegraphics[width=174mm]{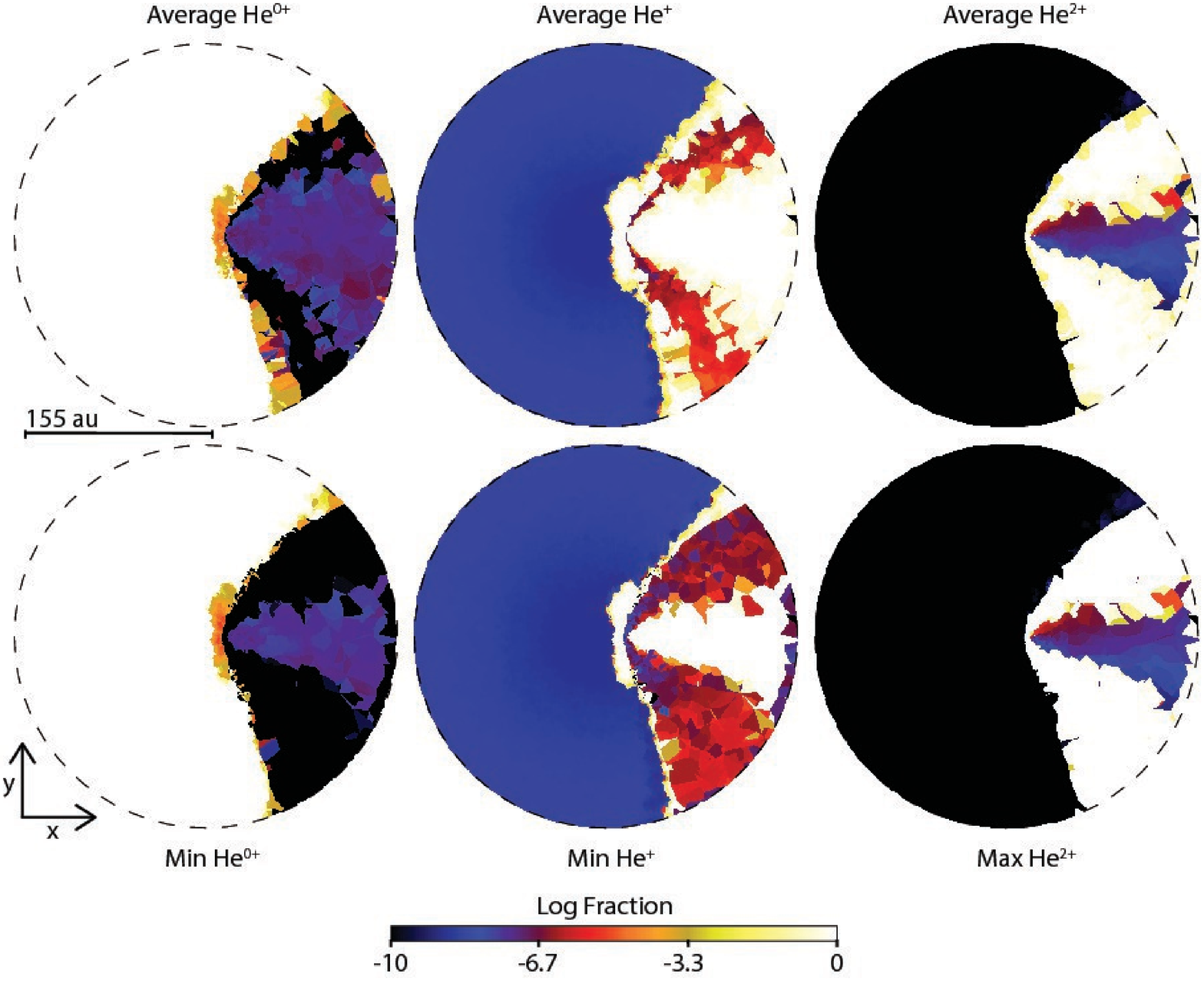}}
   \caption{Slice in the $xy$ orbital plane through the 3D simulation volume for the Case~A simulation at apastron. Columns show, from left to right, the computed fractions of \ion{He}{0+}, \ion{He}{+} and \ion{He}{2+} (log scale). Images in the top row were computed using the volume average approach (see Section~\ref{ssec:VisIt}). Images in the bottom row show, from left to right, the minimum \ion{He}{0+}, minimum \ion{He}{+} and maximum \ion{He}{2+} values and extents. In this and future plots, the dashed circle marks the edge of the spherical computational domain.}\label{fig:min_vs_average}
 \end{center}
\end{figure*}

When visualizing the \SimpleX simulation output, we would ideally like to render physical quantities that are centred on the original Voronoi cells that compose our 3D unstructured grid. Unfortunately, the Voronoi cells consist of a series of irregular $n$-sided polygons, which makes their visualization quite complex. Instead, it is much more straightforward to visualize the corresponding Delaunay triangulation. In 3D, the Delaunay cells are tetrahedra, which can be visualized using standard visualization tools such as \textsc{visit}\footnote{https://wci.llnl.gov/simulation/computer-codes/visit}. Since the Delaunay cells are tetrahedra, the quantity we visualize is the average of the four vertices that define the tetrahedron cell (i.e. the average of the four Voronoi nuclei). This approach works well for visualizing most physical quantities (e.g. temperature, density, velocity). However, if neighbouring Voronoi nuclei have values which are significantly different (i.e. by several orders of magnitude), this `volume-average' approach may lead to tetrahedral-cell values that are difficult to interpret.

Unfortunately, the fractions of \ion{He}{0+}, \ion{He}{+}, and \ion{He}{2+} can span 10 or more orders of magnitude across the WWIR in \ec. The volume-averaged fractions in the larger grid cells that define the post-shock secondary wind can therefore be difficult to understand, especially when a logarithmic colour scale is used (see Fig.~\ref{fig:min_vs_average}). As an example, consider the fraction of \ion{He}{+} near the contact discontinuity (CD) in the WWIR. If three of the vertices of a tetrahedron cell are highly ionized and have extremely low fractions of \ion{He}{+} ($\lesssim \ten{-10}$), while the fourth vertex has a large fraction of \ion{He}{+} ($\approx 1$), the final \ion{He}{+} fraction visualized over the \emph{entire} cell will be $\sim$0.25. This simply tells us that $\sim 25$\% of the cell's volume is \ion{He}{+}. The problem is that the visualization of the \ion{He}{+} fraction alone tells us nothing about \emph{which} $\sim$25\% of the cell volume is \ion{He}{+}, nor does it tell us directly what percentage of the remaining $\sim$75\% of the cell volume is \ion{He}{0+} or \ion{He}{2+}. Therefore, while correct, the visualized plots of various ionization fractions can be deceiving, since for certain species they give the appearance of physically incorrect locations for the ionization fronts. In our \ion{He}{+} example, there appears to be a thick region of \ion{He}{+} near the CD in the hot, post-shock secondary wind (top-middle panel of Fig.~\ref{fig:min_vs_average}), even though the gas in this region is extremely hot ($\gtrsim$$\ten{6}$~K, see Fig.~\ref{fig:dens_temp_xy}), and should consist entirely of \ion{He}{2+}.

Therefore, instead of the average, we show the minimum vertex value for the fraction of \ion{He}{0+} and the maximum vertex value for the fraction of \ion{He}{2+}. For \ion{He}{+}, we show the maximum value whenever the tetrahedron consists of vertices that are only from the primary wind, and the minimum otherwise (if we were to simply show the minimum of the \ion{He}{+} fraction everywhere, we would underestimate the penetration of \etaB's He-ionizing radiation into \etaA's pre-shock wind). This choice for visualizing our simulations shows an \emph{upper} limit to the ionization state of He, in the sense that it shows the maximum extent of the ionization front (bottom row of Fig.~\ref{fig:min_vs_average}). By using this approach, the ionization structure of He much better follows the temperature structure of the gas (Fig.~\ref{fig:dens_temp_xy}) in places where collisional ionization dominates. Therefore, the physics in our simulations is more truthfully represented. We emphasize that the visualizations are merely to help guide the reader, and neither the physics nor the conclusions of our work depend on them.


\subsection{Application to \ec}\label{ssec:EtaCar}

We focus on the ionization of He at an orbital phase of apastron assuming the same abundance by number of He relative to H as \citet{Hillier_etA_2001}, $n_{\mathrm{He}}/n_{\mathrm{H}} = 0.2$. We employ a single photoionizing source located at the position of \etaB. Details on the nature and implementation of the \etaB spectrum are described in Section~\ref{ssec:Source}.

Collisional ionization equilibrium of the SPH simulation snapshot is used as an initial condition for the \SimpleX simulations. The SPH output is post-processed with \SimpleX until the ionization state reaches an equilibrium value (this typically happens within $\sim$1--2~months of simulation time). We use a simulation time-step of $\sim$5~min, which is sufficiently small for accurate RT calculations of the ionization volumes and fractions.

For simplicity, we neglect the influence of the WWIR X-rays on the He ionization structure at times around apastron since they are highly inefficient at ionizing He \citepalias{Clementel_etA_2014}.


\subsubsection{Influence of \etaA}\label{ssec:Primary}

\begin{figure}
 \begin{center}
   \rmf{\includegraphics[width=84mm]{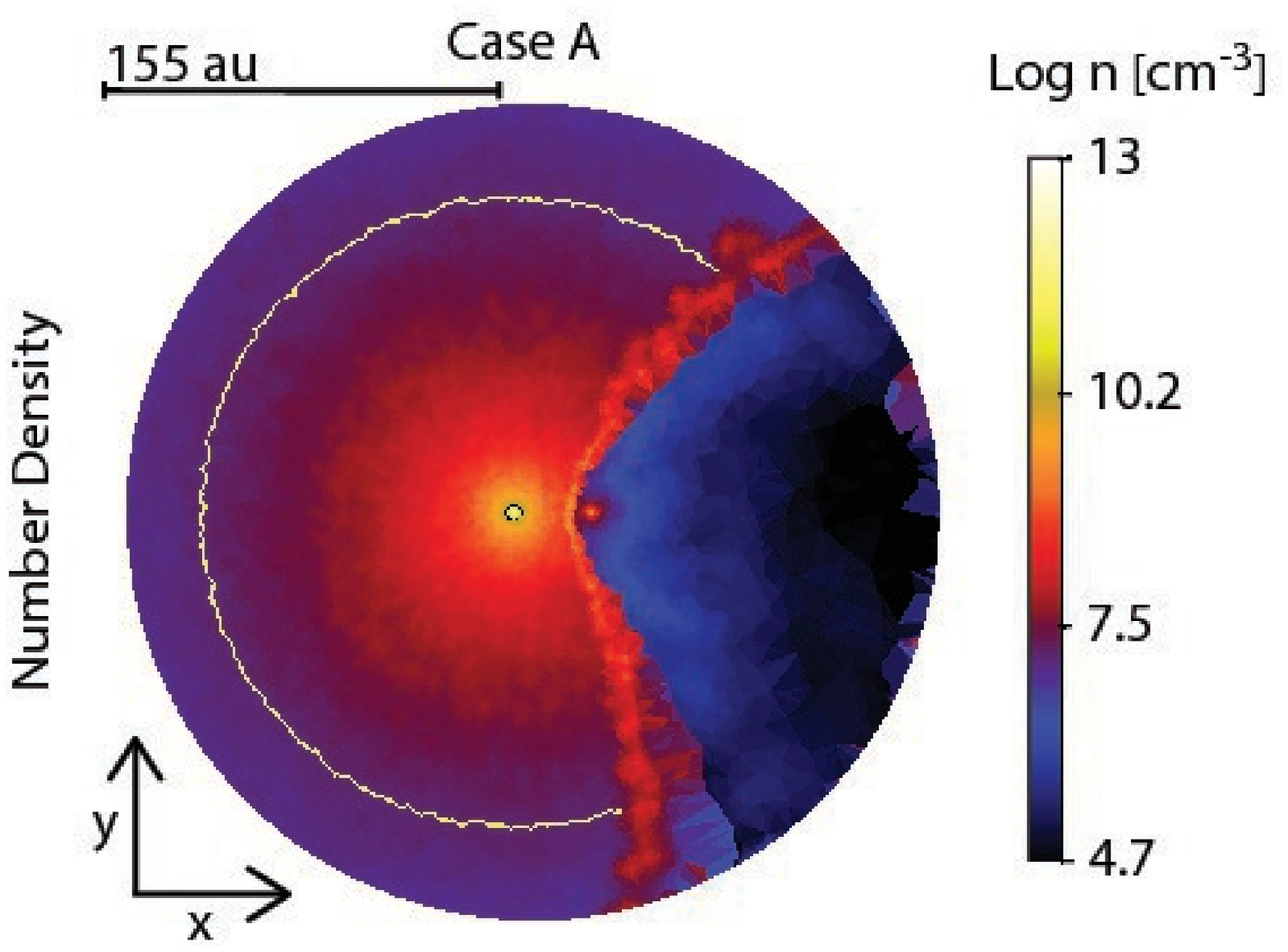}}
   \caption{Slice in the $xy$ orbital plane through the 3D simulation volume for the Case~A simulation at apastron. The black ($r = 3$~au) and yellow ($r = 120$~au) lines indicate the \ion{He}{+} and \ion{H}{+} ionization radii in the pre-shock \etaA wind for simulation Case~A, based on the \textsc{cmfgen} models of \citet{Hillier_etA_2001, Hillier_etA_2006} and \citet{Groh_etA_2012a}. Radii for Case~B are only marginally larger than shown. For Case~C, the \ion{He}{+} radius is comparable to the yellow line (see Fig.~\ref{fig:ion_He_xy}), while H is \ion{H}{+} throughout the entire simulation domain.}\label{fig:etaA_ion_radius}
 \end{center}
\end{figure}

As in \citetalias{Clementel_etA_2014}, we neglect the \etaA ionizing source for simplicity. We also do not consider the ionization structure of hydrogen (although it is included in our calculations). In this work, we focus on the influence of \etaB's He-ionizing radiation on the WWIRs and \etaA's pre-shock wind. \citet{Hillier_etA_2001, Hillier_etA_2006} and \citet{Groh_etA_2012a} fitted the optical and UV spectra of \etaA and computed the ionization structure of H and He within the optically-thick wind of \etaA for different \MdotA. They found that the \ion{H}{+} region around \etaA extends radially $\sim$120--125~au for simulation Cases~A ($\MdotA = \sci{8.5}{-4}$~\Msy; see e.g. the yellow line in Fig.~\ref{fig:etaA_ion_radius}) and B ($\MdotA = \sci{4.8}{-4}$~\Msy). For Case~C ($\MdotA = \sci{2.4}{-4}$~\Msy), hydrogen is fully ionized throughout the entire \etaA wind ($r \gtrsim 5500$~au).

Due to the smaller domain size of the simulations used in this paper ($r = 155$~au), and because \etaB will fully ionize \ion{H}{0+} throughout its wind, little information is gained by examining the ionization structure of hydrogen in our simulations. In principle, we might gain information about the ionization state of H in the cold, post-shock \etaA gas, but this depends strongly on the population of the $n = 2$ state of \ion{H}{0+} in this region and the \etaA ionizing source. Lower energy (10.2~eV) photons could populate the $n = 2$ state of any \ion{H}{0+} in the post-shock \etaA gas, which could then be ionized to \ion{H}{+} by 3.4~eV photons. Thus, it is possible that H is ionized everywhere in the inner $\sim$120--150~au region around \etaA, including the WWIRs.

In contrast, He is mainly neutral in the wind of \etaA. As shown in Fig.~\ref{fig:etaA_ion_radius} (black line), the extent of the \ion{He}{+} region in \etaA's wind, for Cases~A ($r \sim$~3~au) and B ($r \sim$~7.5~au), is much smaller \citep{Hillier_etA_2001, Hillier_etA_2006, Groh_etA_2012a}. The energy level structure of He is such that high-energy photons ($\sim$19.8~eV) are needed to populate even the lowest excited states \citep{Nielsen_etA_2007}. \etaB is the only known source of such photons in \ec. \HeI lines are thought to represent the highly excited regions of \etaA's wind and/or the WWIR \citep{Nielsen_etA_2007}. It may be the case that \ion{He}{0+}-ionizing photons from \etaB are able to penetrate the WWIRs and reach the \ion{He}{+} region deep within \etaA's wind. The inner \etaA~\ion{He}{+} region would be effectively transparent to such photons, which may allow them to pass through to the \ion{He}{0+} zone on the back side of \etaA (the side facing away from \etaB). Thus, determining the correct overall He ionization structure requires some method of mimicking the internal ionization structure of He in \etaA's pre-shock wind due solely to \etaA.

The simplest method, which we employ for this work, is to set as an initial condition to the \SimpleX simulations the ionization structure of \ion{He}{+} in \etaA's inner wind. To do this, we set the temperature in the innermost \etaA wind to 50,000~K, which is hot enough to singly-ionize \ion{He}{0+} to \ion{He}{+}, but not \ion{He}{2+}. The outer radius of this inner \ion{He}{+} region for each simulation case is set to the appropriate value based on the 1D \textsc{cmfgen} models of \ec by \citet{Hillier_etA_2001, Hillier_etA_2006}, namely, 3~au, 7.5~au, and 120~au for Cases~A, B, and C, respectively (see Fig.~\ref{fig:ion_He_xy}). We do not include the inner \ion{He}{2+} zone in \etaA's wind since in most cases it is of negligible size ($r < 1$~au), and because \etaB produces essentially zero \ion{He}{+}-ionizing photons.


\section{Results}\label{sec:Results}

\begin{figure*}
 \begin{center}
    \rmf{\includegraphics[width=174mm]{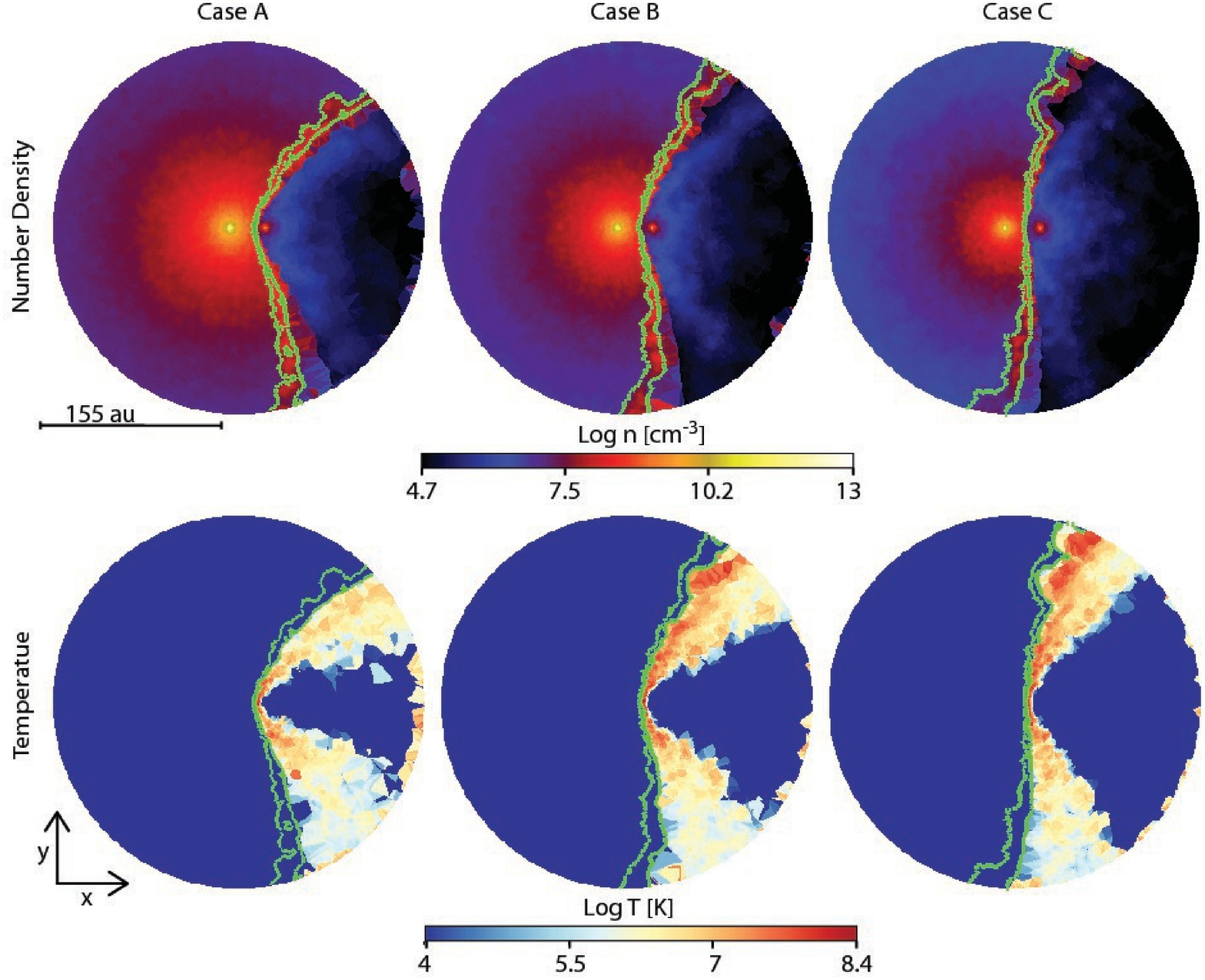}}
    \caption{Slices in the orbital plane through the 3D \SimpleX simulation volume for the three different assumed \MdotA (columns, left to right; Case~A $= \sci{8.5}{-4}$~\Msy, Case~B $= \sci{4.8}{-4}$~\Msy, and Case~C $= \sci{2.4}{-4}$~\Msy). Top row shows the \SimpleX number density (log scale, cgs units), while the bottom row shows the temperature (log scale, K). In this and future plots, the green contour highlights the location of the cold, dense, post-shock primary wind region.}\label{fig:dens_temp_xy}
 \end{center}
\end{figure*}

\begin{figure*}
 \begin{center}
    \rmf{\includegraphics[width=174mm]{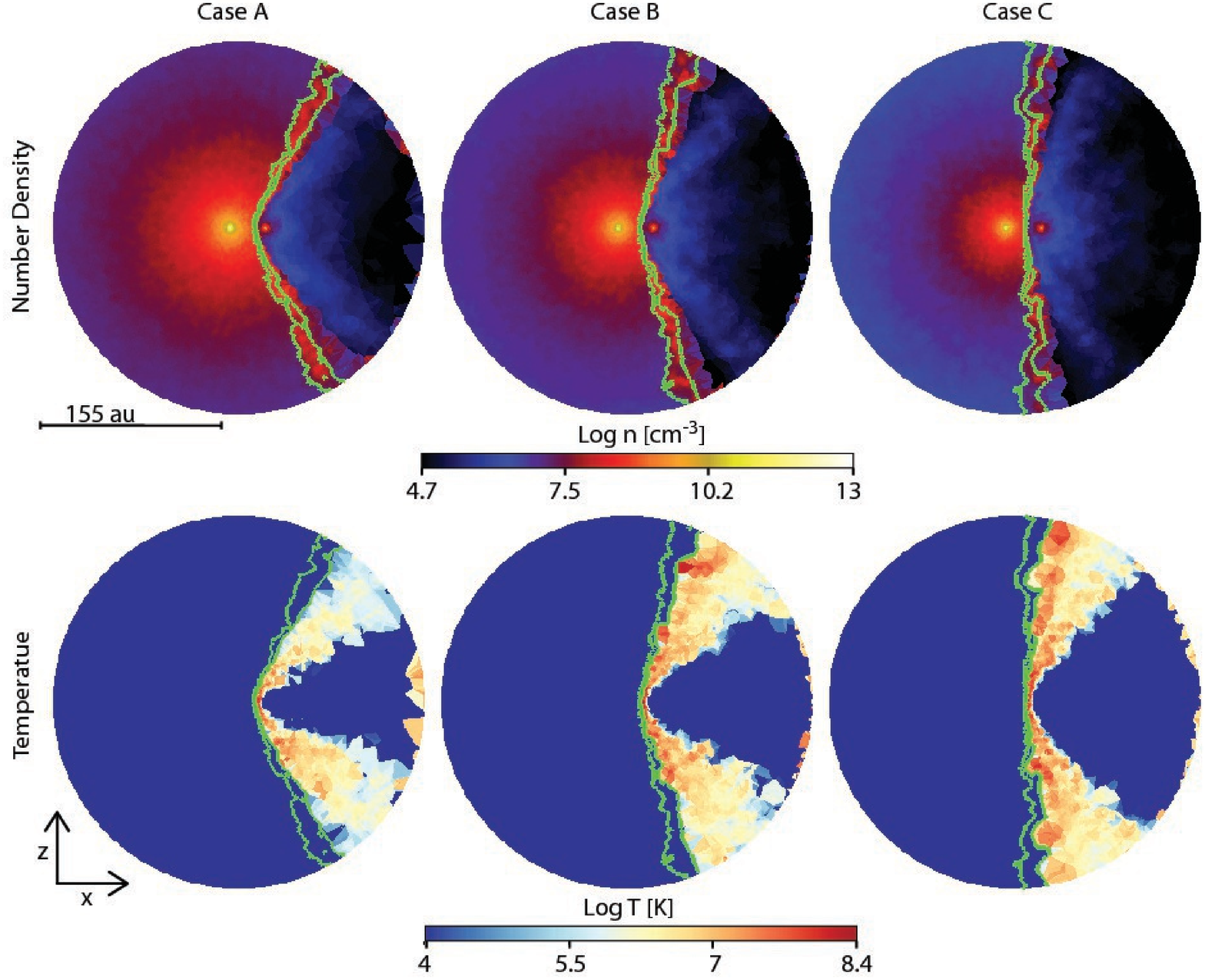}}
    \caption{Same as Fig.~\ref{fig:dens_temp_xy}, but for slices centred in the $xz$ plane.}\label{fig:dens_temp_xz}
 \end{center}
\end{figure*}

Figs.~\ref{fig:dens_temp_xy} and \ref{fig:dens_temp_xz} display the number density and temperature in the orbital and $xz$ planes for simulation Cases~A--C. Due to the smaller spatial size of these simulations, the density and temperature structures are less complex than those described in \citetalias{Clementel_etA_2014}. As shown by \citet{Okazaki_etA_2008, Parkin_etA_2011, Madura_Groh_2012, Madura_etA_2012, Madura_etA_2013}, the lower density faster \etaB wind carves a large cavity out of the slower, denser wind of \etaA for the majority of the orbital period. Around apastron, this cavity and the WWIR have an nearly axisymmetric conical shape, with the opening angle increasing as the value of \MdotA decreases. The apex of the WWIR also moves closer to \etaA as \MdotA is lowered, due to the change in wind momentum balance. The WWIR consists, on the secondary side of the CD (right-hand side of the panels), of a distended shock containing hot ($T \gtrsim$~\ten{6}--\ten{8}~K) low-density \etaB wind material. On the primary side of the CD, the post-shock \etaA gas (green contours) is much thinner and colder ($T \approx \ten{4}$~K).

The bottom row of Figs.~\ref{fig:dens_temp_xy} and \ref{fig:dens_temp_xz} show that, in the orbital plane, the overall fraction of \etaB's wind that is shock heated increases, relative to the total area of secondary wind, with \MdotA. Additionally, the amount of the hottest gas (in red) located in the arms of the WWIR increases as \MdotA decreases. This happens because, for a given pre-shock wind speed, more oblique shocks (i.e. Case~A versus Cases~B and C) produce lower post-shock temperatures (\citealt{Pittard_2009}; \citetalias{Madura_etA_2013}). For all three \MdotA, the hottest gas is located at the apex of the WWIR. However, there is an asymmetry in the temperature of the post-shock \etaB gas in the arms of the WWIR, with the gas in the leading arm hotter than the gas in the trailing arm. This is a result of the different pre-shock wind speeds in the two arms, caused by the orbital motion. The wind in the direction of orbital motion has an additional component added to its velocity, due to the velocity of the star about the system centre of mass. In the opposite direction of orbital motion, this component is subtracted, so that the wind is slightly slower in that direction, and therefore the shock slightly cooler. The post-shock \etaA wind region appears to become slightly thinner and less dense the lower the value of \MdotA, and the WWIR seems to become more unstable. However, standard SPH schemes are known for under resolving certain hydrodynamic instabilities \citep{Agertz_etA_2007, Price_2008}, so these results should be interpreted with caution. For further details on the density and temperature structures of the winds and the effects of different \MdotA, see \citetalias{Madura_etA_2013}.

To help provide a scale comparison with the larger domain simulations of \citetalias{Clementel_etA_2014}, we point out that the remnant of the expanding shell of \etaA wind created during the previous periastron passage (as described in \citetalias{Madura_etA_2013}) is visible at the outer edge of the simulation domain on the apastron side of the system (see the shock-heated gas in the bottom row of Figs.~\ref{fig:dens_temp_xy} and \ref{fig:dens_temp_xz}). Furthermore, to assist the reader in interpreting the He ionization plots, we have outlined the location of the cold, dense, post-shock \etaA wind in Figs.~\ref{fig:dens_temp_xy}--\ref{fig:ion_He_xz} using a green contour. Finally, we note that, as illustrated in the rightmost column of Fig.~\ref{fig:dens_mesh}, on this simulation scale, slices in the $yz$ plane through the system centre of mass only sample the pre-shock wind of \etaA. They thus provide little new relevant information on the ionization structure of He at apastron. Hence, we focus our discussion on the results in the orbital ($xy$, Fig.~\ref{fig:ion_He_xy}) and $xz$ (Fig.~\ref{fig:ion_He_xz}) planes.


\subsection{Overall He Ionization Structure and Influence of \MdotA}\label{ssec:Ion&Mdot}

\subsubsection{The orbital plane}

\begin{figure*}
 \begin{center}
    \rmf{\includegraphics[width=174mm]{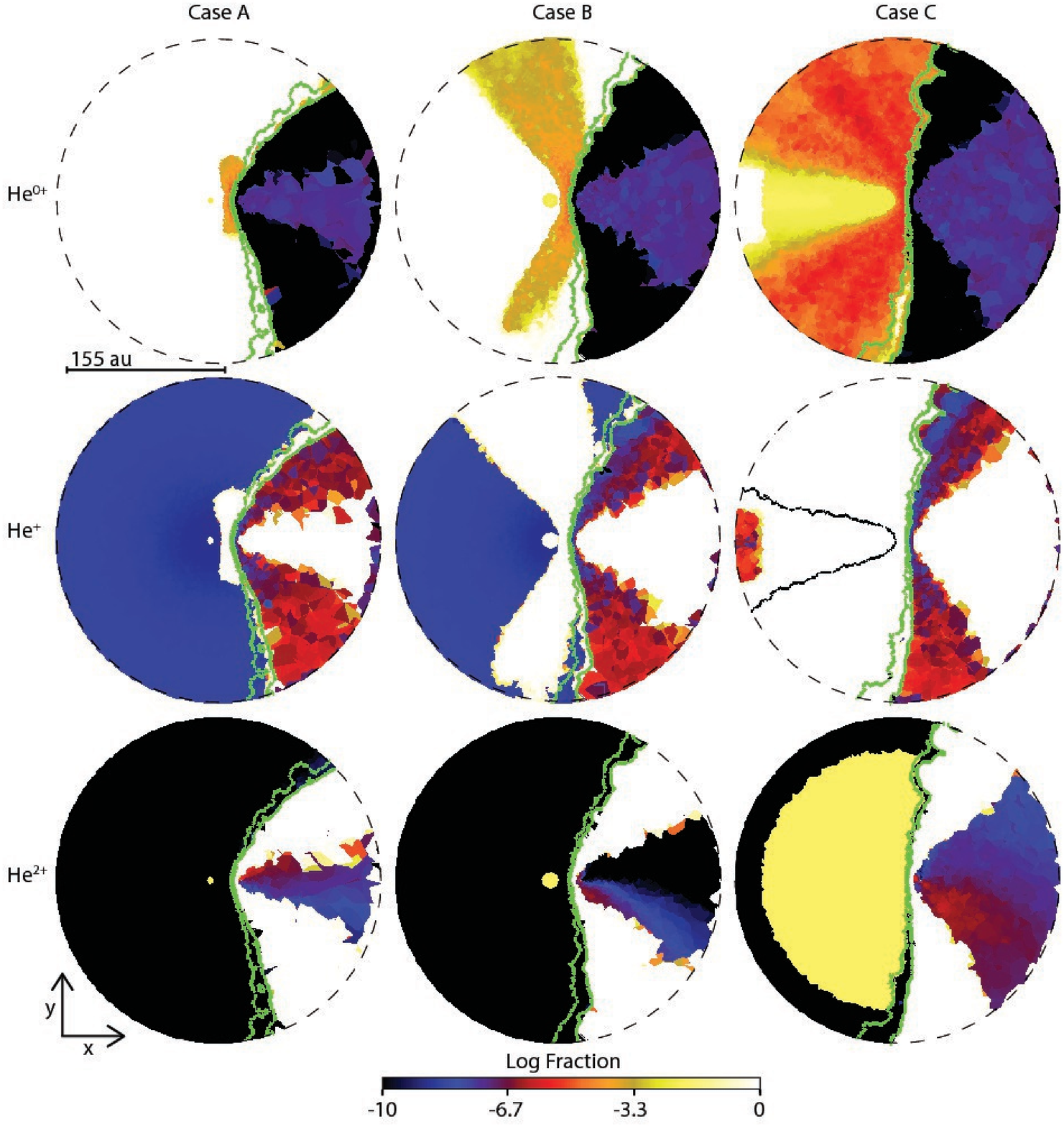}}
    \caption{Slices in the orbital plane through the 3D \SimpleX simulation volume for the three different assumed \MdotA (columns, left to right). Rows show, from top to bottom, the computed fractions of \ion{He}{0+}, \ion{He}{+}, and \ion{He}{2+} (log scale). The circular ionization structure around \etaA (i.e. the yellow circular areas in the \ion{He}{0+} and \ion{He}{2+} rows) is the region where we have set He to \ion{He}{+} in the pre-shock primary wind (see Section~\ref{ssec:Primary}). The black contour in the right-hand panel of the middle row (Case~C, \ion{He}{+}) marks the location of the \ion{He}{+} ionization front when the ionization structure of \etaA's inner wind is excluded in the calculations.}\label{fig:ion_He_xy}
 \end{center}
\end{figure*}

Fig.~\ref{fig:ion_He_xy} illustrates the fractions of \ion{He}{0+}, \ion{He}{+} and \ion{He}{2+} (rows, top to bottom) in the orbital plane for the three \MdotA simulations (Cases~A--C, from left to right). In all three cases, on the secondary side of the WWIR, the ionization state of the high temperature shock-heated gas is dominated by collisional ionization. At such high temperatures ($T \gtrsim \ten{6}$~K) helium is fully-ionized to \ion{He}{2+} (white area in the bottom row). The unperturbed expanding secondary wind located between the two arms of the WWIR is, instead, principally composed of \ion{He}{+} due to photoionization by \etaB. This is an expected difference, compared to the results in \citetalias{Clementel_etA_2014}, connected with the better approximation of the \etaB spectrum using three frequency bins. As discussed in Section~\ref{ssec:Primary}, \etaB should not produce many \ion{He}{+}-ionizing photons. Therefore, the low fraction of \ion{He}{2+} in its unshocked wind is expected. Note also that there is no \ion{He}{2+} in the pre- and post-shock primary wind, as expected. \ion{He}{2+} thus appears to be an excellent tracer of the hot, post-shock \etaB wind region. One important consequence of the larger opening angle of the WWIR for lower \MdotA is an increase in the volume of the \ion{He}{+} region in \etaB's wind on the apastron side of the system, and an increase in the angle between the two arms of \ion{He}{2+}.

While the situation on the \etaB-side of the system does not depend strongly on \MdotA, the He structures on the periastron (\etaA) side of the system are quite different for the three \MdotA. Our simulations show that the \ion{He}{0+}-ionizing photons are able to penetrate into the unperturbed primary wind to varying degrees for all \MdotA, but the detailed results depend drastically on \MdotA. In Case~A, the \ion{He}{+} structure is much smaller and closer to the WWIR apex than in Cases~B--C. Aside from this central area around the apex of the WWIR, the \ion{He}{+} front appears to extend only about half way into the post-shock primary wind. \emph{Most importantly, there is no layer of \ion{He}{+} in the pre-shock \etaA wind that borders the entire WWIR}. This is in contrast to the simple models proposed by e.g. \citet{Martin_etA_2006, Humphreys_etA_2008}, and \citet{Mehner_etA_2012}. The cold, dense, post-shock \etaA wind, in our simulation, absorbs most of the \ion{He}{0+}-ionizing photons from \etaB.

The middle row of Fig.~\ref{fig:ion_He_xy} shows that a decrease in \MdotA leads to a deeper penetration of the \ion{He}{0+}-ionizing photons into the primary wind and, consequently, a much larger volume of \ion{He}{+} on the primary side of the system. \etaB is able to effectively ionize a significant volume of the primary wind in Case~B (the edge of the ionization front is just visible at the bottom edge of the panel in the pre-shock primary wind), while in Case~C, \etaB is able to ionize He in nearly the entire pre- and post-shock primary wind (on this simulation domain scale). Note also the `bent wing' geometry of the \ion{He}{+} ionization front that penetrates into \etaA's pre-shock wind in Case~B, which is caused by the radial dependence of the density in \etaA's wind.

The introduction of the inner \ion{He}{+} region (due to \etaA photoionization, see Section~\ref{ssec:Primary}) has no noticeable effect on the He ionization structure for Cases~A and B (middle row of Fig.~\ref{fig:ion_He_xy}). In both cases, \ion{He}{0+}-ionizing photons from \etaB are unable to reach the deepest parts of the pre-shock primary wind. On the other hand, the much larger ionized sphere in Case~C ($r = 120$~au) further increases the volume of \ion{He}{+} in \etaA's wind. The black line, in the right-hand panel of the middle row of Fig.~\ref{fig:ion_He_xy}, marks the area that is \ion{He}{0+} when the ionization due to \etaA is excluded from our calculations. As one would expect, when the influence of \etaA is neglected, as \MdotA decreases, a larger volume is ionized. Only a narrow column focused directly behind and away from \etaA remains neutral.

In Cases~A and B, there is also a noticeable asymmetry in ionization between the leading and trailing arms of the WWIR. The leading arm shows a higher (in value) and wider (in spatial extent) presence of \ion{He}{+}, both in the post-shock primary wind and the unperturbed primary wind. A possible explanation for this difference might be found in the slight differences in temperature and density between the two arms. Since the leading arm of the WWIR has higher gas temperatures, \ion{He}{0+}-ionizing photons may be able to more easily penetrate the post-shock \etaB gas in the leading arm, causing more ionization of the pre- and post-shock \etaA wind.

\subsubsection{The $xz$ plane}

\begin{figure*}
 \begin{center}
    \rmf{\includegraphics[width=174mm]{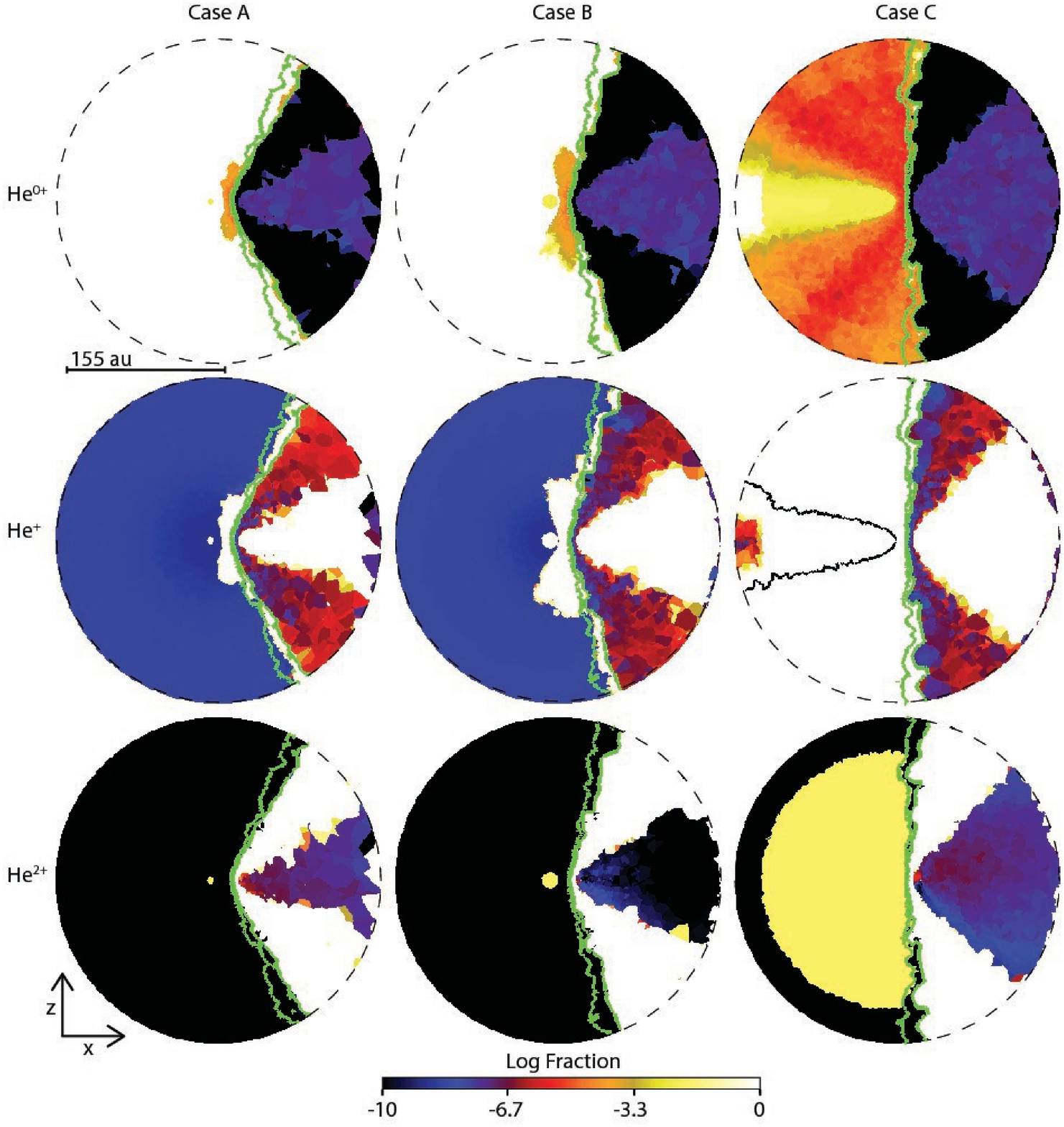}}
    \caption{Same as Fig.~\ref{fig:ion_He_xy}, but for slices centred in the $xz$ plane.}\label{fig:ion_He_xz}
 \end{center}
\end{figure*}

Fig.~\ref{fig:ion_He_xz} shows the fractions of \ion{He}{0+}, \ion{He}{+}, and \ion{He}{2+} in the $xz$ plane for Cases~A--C. The ionization structure in this plane exhibits the same trends, as a function of \MdotA, as the orbital plane. This is expected due to the nearly axisymmetric nature of the WWIR around apastron. However, there is one striking difference, namely, the structure of \ion{He}{+} in the Case~B simulation. In this case, the \ion{He}{+} structures on the primary side of the CD are much smaller in the $xz$ plane than those in the orbital plane. We speculate that this may be due to the less turbulent nature of the WWIR in the plane perpendicular to the orbital motion. Instabilities in the WWIR may be more prevalent in the orbital plane, causing gaps to arise in the post-shock \etaA wind that allow \ion{He}{0+}-ionizing photons from \etaB to more easily penetrate into the pre-shock \etaA wind. However, detailed studies of the 3D structure of complex WWIRs (like those in \ec) in regions above and below the orbital plane, and how various instabilities affect this structure, currently do not exist in the literature. Future detailed simulations of such 3D WWIRs are necessary to determine if our interpretation is correct.

\section{Discussion}\label{sec:Discussion}

Our \SimpleX results show that, assuming for \etaB a typical ionizing flux appropriate for an O-type star with $\mathrm{T_{eff}} \approx 40,000$~K, \etaB singly-ionizes He throughout its unshocked wind, as expected. Moreover, extremely high temperatures ($\gtrsim$$\ten{6}$~K) cause He to be doubly-ionized in the post-shock \etaB gas. Our results also help rule out very low values of \MdotA approaching \sci{2.4}{-4}~\Msy. Our Case~C simulations show that for such low \MdotA, \etaB is able to singly-ionize He throughout practically our entire computational domain, including the dense pre- and post-shock primary wind regions. The ionized volume is even bigger if one includes the He ionization structure of \etaA's inner wind, wherein \etaA already singly-ionizes He out to a radius of $\sim$120~au in its wind \citep{Hillier_etA_2001, Hillier_etA_2006}. This makes it easier for ionizing photons from \etaB to penetrate \etaA's wind and enlarge the He ionization zone. In such a scenario, one would expect a significant amount of \HeII $\lambda$4686 emission from the dense, ionized primary wind and WWIR, even around apastron, which is not observed \citep{Hillier_etA_2006, Teodoro_etA_2012}. The strength of the emission in the broad \HeI lines is also expected to be much stronger than observed if \MdotA were so low \citep{Hillier_etA_2006}.

A similar situation is envisioned in Case~B, namely, stronger \HeI and \HeII emission lines compared to Case~A. This is true even though the inner \ion{He}{+} ionization radius produced by \etaA in its wind is much smaller than in Case~C. However, \ion{He}{0+}-ionizing photons are still able to penetrate the dense WWIR and ionize a significant volume of \etaA's pre-shock wind. Determining the strength of such He emission lines in Case~B is beyond the scope of this paper, so we unfortunately cannot rule out at this time that such emission would be in direct disagreement with observations. Still, we would naively expect, based on our results, significant detectable \HeII emission at times around apastron, which is currently not observed. Thus, Case~B (\sci{4.8}{-4}~\Msy) may represent a lower limit on the current value of \MdotA.

The location of the \ion{He}{+} region in our simulations provides further support for binary orientations in which apastron is on our side of the system \citep[e.g.][]{Damineli_etA_2008b, Okazaki_etA_2008, Parkin_etA_2009, Parkin_etA_2011, Madura_etA_2012}. The observed \HeI emissions in \ec are primarily blueshifted along most of the 5.5~yr cycle, and the P~Cygni absorptions (which must be formed on our side of \etaA, the continuum source) are weak for most of the cycle \citep{Hillier_etA_2001, Hillier_etA_2006, Nielsen_etA_2007, Damineli_etA_2008b}. Both of these facts indicate that the side of the system facing us is more ionized than the far side of \etaA's wind. Figs.~\ref{fig:ion_He_xy} and \ref{fig:ion_He_xz} show that this is true only if \etaB, and hence apastron, is on the observer's side of the system for most of the binary orbit.

The \SimpleX results appear to favour a formation scenario for the \HeI lines that is more-or-less consistent with that proposed by e.g. \citet{Nielsen_etA_2007, Damineli_etA_2008b}, with some modifications. In the following, we consider the Case~A simulation results, which most likely represent \etaA's current mass-loss rate \citepalias{Madura_etA_2013}. The simulations clearly show that with apastron on the observer's side of the system, \etaB ionizes portions of the WWIR and pre-shock \etaA wind flowing towards the observer. Thus, for most of the orbit, the \HeI emission lines would form mostly in the blueshifted part of \etaA's wind, consistent with the observations. The double, and sometimes multiple, peak profiles observed in \HeI emission are consistent with emission arising from spatially separated regions in the WWIR. The two ionized dense arms of post-shock \etaA wind (Figs.~\ref{fig:ion_He_xy} and \ref{fig:ion_He_xz}) are very likely the sources of these emission peaks. The broader general component of the \HeI emission likely arises in both the central \etaA wind (due to ionization by \etaA itself) and the larger ionization zone within the pre-shock \etaA wind located near the WWIR apex.

The observed \HeI absorption is also blueshifted over \ec's entire spectroscopic cycle, but is relatively weak around apastron \citep{Nielsen_etA_2007}. Based on our modelling, this is consistent with the idea that the absorption arises in the ionized pre-shock primary wind located near the WWIR apex, and the ionized portions of the WWIR (specifically, the dense post-shock \etaA wind). The gas velocities in these two regions, which are approximately equal to the terminal velocity of \etaA's wind, are consistent with the observed velocity of the P~Cygni absorption in the \HeI lines ($\sim -300$ to $-600$~\kms). Which region dominates the \HeI absorption in line of sight is uncertain at this point, but the much denser (by at least an order of magnitude) WWIR likely dominates, especially if one considers sight lines that are nearly parallel to and intersecting the WWIR surface (which is likely the case since the binary is inclined by $\sim 45^{\circ}$ away from the observer; see \citealt{Madura_etA_2012, Madura_etA_2013}). Thus, contrary to the assumption by \citet{Nielsen_etA_2007}, the observed \HeI absorption is very likely dominated by material in the post-shock \etaA wind, rather than in the pre-shock \etaA wind. This is the scenario favoured by \citet{Damineli_etA_2008b}.

Another important difference between our simulation results and simpler models for \ec's He ionization zones \citep[e.g.][]{Martin_etA_2006, Humphreys_etA_2008, Mehner_etA_2012} is the detailed ionization structure of the gas in the pre-shock \etaA wind that borders the WWIR. This is due mainly to the complex structure of the WWIR, which in toy models is too simplistic. The WWIR consists of a region of hot, compressed (by roughly a factor of 4) post-shock \etaB wind separated by a CD from a thin region of very high density, colder ($T\sim$~\ten{4}~K) post-shock primary wind. The dense, turbulent post-shock \etaA wind absorbs most of \etaB's \ion{He}{0+}-ionizing photons and greatly affects the ionization structure of \etaA's pre-shock wind. Detailed examination of Figs.~\ref{fig:ion_He_xy} and \ref{fig:ion_He_xz} shows that with the exception of a $r\sim$~75--80~au region about the WWIR apex, the \ion{He}{+} ionization front due to \etaB penetrates only approximately half way into the post-shock \etaA wind region. Only in the region near the WWIR apex does \etaB singly-ionize He in \etaA's pre-shock wind. At larger distances from \etaA along the walls of the WWIR, He remains neutral in the pre-shock \etaA wind. Thus, simple models like that shown in fig.~12 of \citealt{Humphreys_etA_2008} and fig.~5a of \citealt{Mehner_etA_2012} are only partially correct\footnote{They appear to neglect the post-shock \etaA wind region and CD.}, and very dependent upon scalelength (\citealt{Martin_etA_2006, Humphreys_etA_2008} and \citealt{Mehner_etA_2012} show only a small region near the stars at apastron). The \ion{He}{+} region around the WWIR apex is also thinner and more geometrically similar to the WWIR in such models than we find in our simulations. The \ion{He}{+} region in the pre-shock \etaA wind in our simulations is more similar to fig.~5b of \citet{Mehner_etA_2012}, but with the wings of \ion{He}{+} not bending back farther than the location of \etaA.

\citet{Mehner_etA_2012} used their fig.~5 to illustrate their interpretation for why the strength of the \HeI P~Cygni absorption observed in \ec has been gradually increasing since 1998, with a sudden absorption increase after the 2009 periastron event \citep{Groh_Daminelli_2004, Mehner_etA_2010, Mehner_etA_2012}. \citet{Mehner_etA_2012} attribute the observed increase in absorption, and other changes, to a gradual decrease of \MdotA by a factor of $\sim$2--3 between 1999 and 2010. They suggest that a drop in \MdotA would change the He ionization structure of \etaA's pre-shock wind, as depicted in their fig.~5. \citet{Mehner_etA_2012} suggest that the enlarging of the \ion{He}{+} zone in \etaA's pre-shock wind with decreasing \MdotA would lead to more \HeI absorption in line of sight.

Our \SimpleX simulations show that, in the orbital plane, the behaviour of the \ion{He}{+} ionization zone in \etaA's pre-shock wind with decreasing \MdotA, as suggested by \citet{Mehner_etA_2012}, is partially correct. We find that a decrease in \MdotA does enlarge the \ion{He}{+} zone in the pre-shock \etaA wind. Moving from Case~A to Case~B, two large `wings' of \ion{He}{+} develop in the pre-shock \etaA wind that bend backward around \etaA. The post-shock \etaA wind also becomes more ionized and contains more \ion{He}{+} out to larger radii from \etaA. However, the \ion{He}{+} wings in the pre-shock \etaA wind do not continuously border the WWIR at all radii in Case~B. Near the outer edges of the simulation in the orbital plane, there are clear gaps between the pre-shock \etaA\ \ion{He}{+} zone and the WWIR. In these small regions, He remains neutral. Moreover, changes to the He ionization structure with decreasing \MdotA are confined mostly to the orbital plane when moving from Case~A to Case~B. The Case~B simulation shows that the \ion{He}{+} ionization zone in the $xz$ plane is much smaller and remains concentrated near the WWIR apex, with no large \ion{He}{+} wings extending back behind \etaA. Decreasing \MdotA even more from Case~B to Case~C results in \etaB ionizing He throughout the entire simulation domain. As discussed above and in \citetalias{Madura_etA_2013}, such a low \MdotA appears to be ruled out by numerous observations.

Based on these results, a small-to-moderate decrease (by no more than a factor of 2) in \MdotA, similar to the simple model suggested by \citet{Mehner_etA_2012}, may help explain the observed increase in \HeI absorption. The key question is, how much \HeI emission and absorption does an ionization structure like that in Case~B produce in line of sight? One might expect such a large change in the \ion{He}{+} zone within \etaA's pre-shock wind to result in a significant detectable change in the amounts of \HeI and \HeII emission. An increase in the amount of \ion{He}{+} near the WWIR that is available to be ionized by very soft X-rays generated in the post-shock \etaA gas as it radiatively cools should lead to increased \HeII $\lambda$4686 emission throughout the binary orbit. Significant changes in the amount of \HeII $\lambda$4686 emission during the broad part of \ec's orbit are not observed though, and the behaviour of the $\lambda$4686 emission during recent spectroscopic events has been amazingly similar (Teodoro et al., in preparation). The lack of any major changes in the observed $\lambda$4686 emission may imply that the amount of \ion{He}{+} available in the pre-shock \etaA wind at phases around apastron is not the key factor in determining the amount of $\lambda$4686 emission. Rather, there may be a limited number of suitable photons generated in the WWIR that can produce $\lambda$4686 emission in the available \ion{He}{+} zone. As such, we suggest that the small amount of \HeII $\lambda$4686 emission that is observed across \ec's entire orbital cycle originates in the \ion{He}{+} zone in \etaA's pre-shock wind that surrounds the apex of the WWIR, with the high-energy photons required for such emission originating in the nearby post-shock \etaA wind.

However, the above scenario does not remedy the problem of why there has been no observed increase in the amount of \HeI emission. If the efficiency for creating \HeII $\lambda$4686 emission around apastron is low, even though the amount of \ion{He}{+} in \etaA's wind is high, one would expect increased \HeI emission. A possible explanation for the lack of increased emission is that the larger volume of ionized post-shock \etaA wind is absorbing the \ion{He}{0+} photons generated in the larger pre-shock \etaA\ \ion{He}{+} zone. The details of how the amounts of emission and absorption in line of sight change as the WWIR opening angle increases (due to the decreased \MdotA and altered wind momentum ratio) are poorly understood at this time, and it is unclear whether any decrease in \MdotA was gradual (between $\sim$1998 and now) or sudden (during/after the 2009 periastron event). More detailed observations and theoretical models are needed to better understand the nature of \ec's \HeI lines.

Finally, based on our results, we speculate on a possible solution to an interesting problem concerning \ion{He}{0+} in \ec, namely, why there are no observed signatures of the broad \HeI emission lines in spectra obtained before 1944 if \ec is a binary containing a hot companion star \citep[for details, see][]{Humphreys_etA_2008}. This is seen as a problem because it is assumed that \etaB, if it exists and is a hot, massive star, should always ionize a significant volume of \etaA's pre-shock stellar wind, leading to detectable \HeI emission. Such emission is only marginally present (at best) in early spectra obtained between \ec's second eruption in the 1890s and 1944. \citet{Humphreys_etA_2008} argue that, even if \MdotA was larger in the past, \etaB would still have ionized \etaA's pre-shock wind, which should lead to \HeI emission in the early spectra.

As shown in Section~\ref{sec:Results} and discussed above, the results of detailed 3D hydrodynamical and RT simulations can differ substantially from the expectations of a simple model like that in fig.~12 of \citet{Humphreys_etA_2008}. Our Case~A simulations show that \etaB's \ion{He}{0+}-ionizing photons only moderately penetrate the dense post-shock \etaA wind at radii $\gtrsim$75~au from the star, and that only the pre-shock wind in the inner $\sim$75~au around the WWIR apex is \ion{He}{+}. We also see from our Case~B and Case~C results that a factor of 2--4 change in \MdotA can lead to major changes in the ionization structure of He in \etaA's pre-shock wind. Therefore, we suggest that the reason broad \HeI lines were not observed in spectra of \ec before 1944 is because \MdotA was larger then (by at least a factor of 2--4, or more) than it is now.

While \SimpleX results using 3D SPH simulations that employ a higher \MdotA are needed to test this hypothesis, we can speculate what the resulting He-ionization structure would look like in such a situation where \MdotA is roughly a factor of 2 or more larger than its present value. Given the results in Figs.~\ref{fig:ion_He_xy} and \ref{fig:ion_He_xz}, the post-shock \etaA wind was probably so dense it absorbed nearly all incoming \ion{He}{0+}-ionizing photons from \etaB, preventing ionization of the \ion{He}{0+} in \etaA's pre-shock wind. There was also probably much less \ion{He}{+} in the post-shock \etaA wind since any such \ion{He}{+} that formed there would likely quickly recombine due to the extremely high densities in that region. There would moreover be little-to-no intrinsic \HeI emission from \etaA itself, since the inner \ion{He}{+} zone in its pre-shock wind would extend to $r<$~2.5~au\footnote{The exact value requires a specific assumed value of \MdotA and detailed spectroscopic modelling with e.g. \textsc{cmfgen}.} from the star. Because the wind momentum ratio would be dominated by \etaA's much denser wind, the opening angle of the WWIR would be much smaller, bringing the two arms of hot, post-shock \etaB wind closer together. Conceivably, the two arms could be so close they practically overlap, in which case the wind of \etaB would be dominated by collisionally-ionized \ion{He}{2+}, with perhaps only a very narrow region of \ion{He}{+} extending behind \etaB (caused by photoionization of the remaining receding pre-shock wind). With practically no \ion{He}{+} present in the inner \ec system, there would be no, or only marginal, \HeI emission detectable, thus explaining the early spectral observations.

An enhanced \MdotA between the 1890s and 1944 is quite possible, and even likely, since \ec had just experienced its second eruption, wherein it ejected $\sim$0.1~\Ms and formed the `Little Homunculus' \citep{Ishibashi_etA_2003, Smith_2005}. The dense, slow-moving near-equatorial circumstellar ejecta known as the Weigelt blobs \citep{Weigelt_Ebersberger_1986} were also ejected around this time. It is therefore not unreasonable to suggest that the LBV primary had a larger mass-loss rate following these events, which slowly decreased to a more stable and `normal' value by $\sim$1944. During this period of higher \MdotA, the ionizing flux from \etaB was probably more confined and could not penetrate the denser WWIR, leading to the absence of any significant \HeI emission.


\section{Summary and Conclusions}\label{sec:Summary}

We investigated the effects of the hot secondary star's ionizing flux on \ec's inner winds and WWIR, focusing on the ionization structure of helium during the spectroscopic high state (i.e. orbital phases around apastron). We used \SimpleX to post-process 3D SPH simulation output of the innermost region of the \ec system. Compared to our previous work, we implemented several changes and improvements in the RT simulations. Assigning to the \SimpleX cells the SPH number density, instead of a Voronoi cell-volume approach, results in density distributions in \SimpleX that are less affected by local differences in the SPH particle distribution. Sampling the \etaB blackbody spectrum with three frequency bins and the use of DCT also lead to more physically realistic results, and hence more precise ionization fractions and front locations for H and He. Below we summarize our most important results.

\begin{enumerate}[leftmargin=*, label=\arabic*.]

\item The inclusion of the \ion{He}{+} ionization volumes around the primary star \etaA does not produce relevant differences in the final ionization results for simulations with high \MdotA (Cases~A and B). The exception is Case~C, wherein \MdotA is so low \etaA fully ionizes H everywhere, and singly-ionizes He out to a radius of $\sim$120~au. \etaB is thus able to completely singly-ionize He throughout the entire computation domain in Case~C. We are therefore able to rule out Case~C as a possible current value for \MdotA.

\item The \SimpleX results show that \etaB's \ion{He}{0+}-ionizing photons are able to penetrate the WWIR and reach the unperturbed \etaA wind to varying degrees, depending on the value of \MdotA. The geometry and extent of the He-ionization structures in the pre- and post-shock \etaA wind depend strongly on \MdotA.

\item \ion{He}{0+} is confined to the pre- and post-shock \etaA wind for higher \MdotA. The volume of \ion{He}{+} in the pre- and post-shock \etaA wind increases as \MdotA decreases. As expected, \etaB singly-ionizes He in its pre-shock wind. \ion{He}{2+} is produced, through collisional ionization, in the hot post-shock \etaB wind.

\item The different extents of the He-ionization fronts into \etaA's wind in both the orbital and $xz$ planes might be caused by differences in the density and temperature state of the WWIR. Small holes or regions of lower density in the unstable post-shock \etaA wind region may also allow \ion{He}{0+}-ionizing photons to penetrate the WWIR and ionize different portions of \etaA's pre-shock wind.

\item The location of the \ion{He}{+} region in our simulations provides further support for binary orientations in which apastron is on our side of the system \citep{Damineli_etA_2008b, Okazaki_etA_2008, Parkin_etA_2009, Parkin_etA_2011, Madura_etA_2012, Madura_etA_2013}. The \SimpleX results favour a formation scenario for the \HeI lines that is mostly consistent with that proposed by \citet{Damineli_etA_2008b}.

\item Based on our results, a small-to-moderate decrease (by no more than a factor of 2) in \MdotA, similar to the simple model suggested by \citet{Mehner_etA_2012}, may help explain the observed increase in \HeI absorption in \ec. However, numerous questions remain regarding this scenario, such as why such a decrease in \MdotA does not also lead to an increase in \HeI emission.

\item We suggest that the small amount of \HeII $\lambda$4686 emission observed across \ec's entire orbital cycle originates in the \ion{He}{+} zone in \etaA's pre-shock wind that surrounds the apex of the WWIR, with the high-energy photons required for such emission originating in the nearby post-shock \etaA wind.

\item Finally, we suggest that broad \HeI lines were not observed in spectra of \ec between its 1890s eruption and 1944 \citep{Humphreys_etA_2008} because \MdotA was larger then (by at least a factor of 2--4, maybe more) than it is now. During this period of higher \MdotA, the ionizing flux from \etaB was probably more confined and could not penetrate the denser WWIR, leading to the absence of any significant \HeI emission.

\end{enumerate}

Our investigation of the He-ionization structure of \ec's inner winds and WWIR helps constrain not only \MdotA, but also the nature of the unseen companion \etaB. We have shown that our results help explain, and are in qualitative agreement with, available observations of \ec's \HeI and \HeII lines. In future work, we plan to use these simulations to generate synthetic spectra for comparison to observational data in order to help place tighter constraints on the binary orientation, \MdotA, and \etaB's luminosity and temperature. Future improvements to \SimpleX will allow us to use specific spectral energy distributions for \etaB generated from detailed spectroscopic modelling with \textsc{cmfgen}. These results, together with the ionization structure of He during the spectroscopic low state around periastron (investigated in a subsequent paper), will help us better understand the numerous observed spectral features that arise in the inner $\sim$150~au of the system. Our numerical work also sets the stage for future efforts to couple \SimpleX with modern 3D SPH and grid-based hydrodynamics codes for even more physically realistic 3D time-dependent radiation-hydrodynamics simulations of \ec and other colliding wind binaries.


\section*{Acknowledgements}

We thank Jose Groh, Noel Richardson, and Vincent Icke for useful discussions and comments. TIM is supported by an appointment to the NASA Postdoctoral Program at the Goddard Space Flight Center, administered by Oak Ridge Associated Universities through a contract with NASA. Support for TRG was through programs \#12013, 12508, 12750, 13054, and 13395, provided by NASA through a grant from the Space Telescope Science Institute, which is operated by the Association of Universities for Research in Astronomy, Inc., under NASA contract NAS 5-26555.

\ifdraft{\nobibliography{biblio}} {\bibliography{biblio}}

\begin{thebibliography}{71}
\expandafter\ifx\csname natexlab\endcsname\relax\def\natexlab#1{#1}\fi

\bibitem[{{Agertz} {et~al}\mbox{.}(2007){Agertz}, {Moore}, {Stadel}, {Potter},
  {Miniati}, {Read}, {Mayer}, {Gawryszczak}, {Kravtsov}, {Nordlund}, {Pearce},
  {Quilis}, {Rudd}, {Springel}, {Stone}, {Tasker}, {Teyssier}, {Wadsley}, \&
  {Walder}}]{Agertz_etA_2007}
{Agertz} O. {et~al.}, 2007, \mnras, 380, 963

\bibitem[{{Clementel} {et~al}\mbox{.}(2014){Clementel}, {Madura}, {Kruip},
  {Icke}, \& {Gull}}]{Clementel_etA_2014}
{Clementel} N., {Madura} T.~I., {Kruip} C.~J.~H., {Icke} V., {Gull} T.~R.,
  2014, \mnras, 443, 2475 (C14)

\bibitem[{{Corcoran}(2005)}]{Corcoran_2005}
{Corcoran} M.~F., 2005, \aj, 129, 2018

\bibitem[{{Corcoran} {et~al}\mbox{.}(2010){Corcoran}, {Hamaguchi}, {Pittard},
  {Russell}, {Owocki}, {Parkin}, \& {Okazaki}}]{Corcoran_etA_2010}
{Corcoran} M.~F., {Hamaguchi} K., {Pittard} J.~M., {Russell} C.~M.~P., {Owocki}
  S.~P., {Parkin} E.~R., {Okazaki} A., 2010, \apj, 725, 1528

\bibitem[{{Damineli}(1996)}]{Damineli_1996}
{Damineli} A., 1996, \apjl, 460, L49

\bibitem[{{Damineli}, {Conti} \& {Lopes}(1997){Damineli}, {Conti}, \&
  {Lopes}}]{Damineli_etA_1997}
{Damineli} A., {Conti} P.~S., {Lopes} D.~F., 1997, \na, 2, 107

\bibitem[{{Damineli} {et~al}\mbox{.}(2008{\natexlab{a}}){Damineli}, {Hillier},
  {Corcoran}, {Stahl}, {Groh}, {Arias}, {Teodoro}, {Morrell}, {Gamen},
  {Gonzalez}, {Leister}, {Levato}, {Levenhagen}, {Grosso}, {Colombo}, \&
  {Wallerstein}}]{Damineli_etA_2008b}
{Damineli} A. {et~al.}, 2008{\natexlab{a}}, \mnras, 386, 2330

\bibitem[{{Damineli} {et~al}\mbox{.}(2008{\natexlab{b}}){Damineli}, {Hillier},
  {Corcoran}, {Stahl}, {Levenhagen}, {Leister}, {Groh}, {Teodoro}, {Albacete
  Colombo}, {Gonzalez}, {Arias}, {Levato}, {Grosso}, {Morrell}, {Gamen},
  {Wallerstein}, \& {Niemela}}]{Damineli_etA_2008a}
{Damineli} A. {et~al.}, 2008{\natexlab{b}}, \mnras, 384, 1649

\bibitem[{{Davidson} \& {Humphreys}(1997)}]{Davidson_Humphreys_1997}
{Davidson} K., {Humphreys} R.~M., 1997, \araa, 35, 1

\bibitem[{{Ferland} {et~al}\mbox{.}(1998){Ferland}, {Korista}, {Verner},
  {Ferguson}, {Kingdon}, \& {Verner}}]{Ferland_etA_1998}
{Ferland} G.~J., {Korista} K.~T., {Verner} D.~A., {Ferguson} J.~W., {Kingdon}
  J.~B., {Verner} E.~M., 1998, \pasp, 110, 761

\bibitem[{{Friedrich} {et~al}\mbox{.}(2012){Friedrich}, {Mellema}, {Iliev}, \&
  {Shapiro}}]{Friedrich_etA_2012}
{Friedrich} M.~M., {Mellema} G., {Iliev} I.~T., {Shapiro} P.~R., 2012, \mnras,
  421, 2232

\bibitem[{{Gayley}, {Owocki} \& {Cranmer}(1997){Gayley}, {Owocki}, \&
  {Cranmer}}]{Gayley_etA_1997}
{Gayley} K.~G., {Owocki} S.~P., {Cranmer} S.~R., 1997, \apj, 475, 786

\bibitem[{{Gomez} {et~al}\mbox{.}(2010){Gomez}, {Vlahakis}, {Stretch}, {Dunne},
  {Eales}, {Beelen}, {Gomez}, \& {Edmunds}}]{Gomez_etA_2010}
{Gomez} H.~L., {Vlahakis} C., {Stretch} C.~M., {Dunne} L., {Eales} S.~A.,
  {Beelen} A., {Gomez} E.~L., {Edmunds} M.~G., 2010, \mnras, 401, L48

\bibitem[{{Groh} \& {Damineli}(2004)}]{Groh_Daminelli_2004}
{Groh} J.~H., {Damineli} A., 2004, \ibvs, 5492, 1

\bibitem[{{Groh} {et~al}\mbox{.}(2012{\natexlab{a}}){Groh}, {Hillier},
  {Madura}, \& {Weigelt}}]{Groh_etA_2012a}
{Groh} J.~H., {Hillier} D.~J., {Madura} T.~I., {Weigelt} G.,
  2012{\natexlab{a}}, \mnras, 423, 1623

\bibitem[{{Groh} {et~al}\mbox{.}(2012{\natexlab{b}}){Groh}, {Madura},
  {Hillier}, {Kruip}, \& {Weigelt}}]{Groh_etA_2012b}
{Groh} J.~H., {Madura} T.~I., {Hillier} D.~J., {Kruip} C.~J.~H., {Weigelt} G.,
  2012{\natexlab{b}}, \apjl, 759, L2

\bibitem[{{Groh} {et~al}\mbox{.}(2010{\natexlab{a}}){Groh}, {Madura}, {Owocki},
  {Hillier}, \& {Weigelt}}]{Groh_etA_2010a}
{Groh} J.~H., {Madura} T.~I., {Owocki} S.~P., {Hillier} D.~J., {Weigelt} G.,
  2010{\natexlab{a}}, \apjl, 716, L223

\bibitem[{{Groh} {et~al}\mbox{.}(2010{\natexlab{b}}){Groh}, {Nielsen},
  {Damineli}, {Gull}, {Madura}, {Hillier}, {Teodoro}, {Driebe}, {Weigelt},
  {Hartman}, {Kerber}, {Okazaki}, {Owocki}, {Millour}, {Murakawa}, {Kraus},
  {Hofmann}, \& {Schertl}}]{Groh_etA_2010b}
{Groh} J.~H. {et~al.}, 2010{\natexlab{b}}, \aap, 517, A9

\bibitem[{{Gull} {et~al}\mbox{.}(2011){Gull}, {Madura}, {Groh}, \&
  {Corcoran}}]{Gull_etA_2011}
{Gull} T.~R., {Madura} T.~I., {Groh} J.~H., {Corcoran} M.~F., 2011, \apjl, 743,
  L3

\bibitem[{{Gull} {et~al}\mbox{.}(2009){Gull}, {Nielsen}, {Corcoran}, {Madura},
  {Owocki}, {Russell}, {Hillier}, {Hamaguchi}, {Kober}, {Weis}, {Stahl}, \&
  {Okazaki}}]{Gull_etA_2009}
{Gull} T.~R. {et~al.}, 2009, \mnras, 396, 1308

\bibitem[{{Hamaguchi} {et~al}\mbox{.}(2007){Hamaguchi}, {Corcoran}, {Gull},
  {Ishibashi}, {Pittard}, {Hillier}, {Damineli}, {Davidson}, {Nielsen}, \&
  {Kober}}]{Hamaguchi_etA_2007}
{Hamaguchi} K. {et~al.}, 2007, \apj, 663, 522

\bibitem[{{Hamaguchi} {et~al}\mbox{.}(2014){Hamaguchi}, {Corcoran}, {Russell},
  {Pollock}, {Gull}, {Teodoro}, {Madura}, {Damineli}, \&
  {Pittard}}]{Hamaguchi_etA_2014}
{Hamaguchi} K. {et~al.}, 2014, \apj, 784, 125

\bibitem[{{Henley} {et~al}\mbox{.}(2008){Henley}, {Corcoran}, {Pittard},
  {Stevens}, {Hamaguchi}, \& {Gull}}]{Henley_etA_2008}
{Henley} D.~B., {Corcoran} M.~F., {Pittard} J.~M., {Stevens} I.~R., {Hamaguchi}
  K., {Gull} T.~R., 2008, \apj, 680, 705

\bibitem[{{Hillier} {et~al}\mbox{.}(2001){Hillier}, {Davidson}, {Ishibashi}, \&
  {Gull}}]{Hillier_etA_2001}
{Hillier} D.~J., {Davidson} K., {Ishibashi} K., {Gull} T., 2001, \apj, 553, 837

\bibitem[{{Hillier} {et~al}\mbox{.}(2006){Hillier}, {Gull}, {Nielsen},
  {Sonneborn}, {Iping}, {Smith}, {Corcoran}, {Damineli}, {Hamann}, {Martin}, \&
  {Weis}}]{Hillier_etA_2006}
{Hillier} D.~J. {et~al.}, 2006, \apj, 642, 1098

\bibitem[{{Humphreys}, {Davidson} \& {Koppelman}(2008){Humphreys}, {Davidson},
  \& {Koppelman}}]{Humphreys_etA_2008}
{Humphreys} R.~M., {Davidson} K., {Koppelman} M., 2008, \aj, 135, 1249

\bibitem[{{Ishibashi} {et~al}\mbox{.}(2003){Ishibashi}, {Gull}, {Davidson},
  {Smith}, {Lanz}, {Lindler}, {Feggans}, {Verner}, {Woodgate}, {Kimble},
  {Bowers}, {Kraemer}, {Heap}, {Danks}, {Maran}, {Joseph}, {Kaiser}, {Linsky},
  {Roesler}, \& {Weistrop}}]{Ishibashi_etA_2003}
{Ishibashi} K. {et~al.}, 2003, \aj, 125, 3222

\bibitem[{{Kruip}(2011)}]{Kruip_2011}
{Kruip} C., 2011, PhD thesis, University of Leiden, Leiden, the Netherlands

\bibitem[{{Kruip} {et~al}\mbox{.}(2010){Kruip}, {Paardekooper}, {Clauwens}, \&
  {Icke}}]{Kruip_etA_2010}
{Kruip} C.~J.~H., {Paardekooper} J.-P., {Clauwens} B.~J.~F., {Icke} V., 2010,
  \aap, 515, A78

\bibitem[{{Luo}, {McCray} \& {Mac Low}(1990){Luo}, {McCray}, \& {Mac
  Low}}]{Luo_etA_1990}
{Luo} D., {McCray} R., {Mac Low} M.-M., 1990, \apj, 362, 267

\bibitem[{{Madura}(2010)}]{Madura_2010}
{Madura} T.~I., 2010, PhD thesis, University of Delaware

\bibitem[{{Madura} \& {Groh}(2012)}]{Madura_Groh_2012}
{Madura} T.~I., {Groh} J.~H., 2012, \apjl, 746, L18

\bibitem[{{Madura} {et~al}\mbox{.}(2013){Madura}, {Gull}, {Okazaki}, {Russell},
  {Owocki}, {Groh}, {Corcoran}, {Hamaguchi}, \& {Teodoro}}]{Madura_etA_2013}
{Madura} T.~I. {et~al.}, 2013, \mnras, 436, 3820 (M13)

\bibitem[{{Madura} {et~al}\mbox{.}(2012){Madura}, {Gull}, {Owocki}, {Groh},
  {Okazaki}, \& {Russell}}]{Madura_etA_2012}
{Madura} T.~I., {Gull} T.~R., {Owocki} S.~P., {Groh} J.~H., {Okazaki} A.~T.,
  {Russell} C.~M.~P., 2012, \mnras, 420, 2064

\bibitem[{{Martin} {et~al}\mbox{.}(2006){Martin}, {Davidson}, {Humphreys},
  {Hillier}, \& {Ishibashi}}]{Martin_etA_2006}
{Martin} J.~C., {Davidson} K., {Humphreys} R.~M., {Hillier} D.~J., {Ishibashi}
  K., 2006, \apj, 640, 474

\bibitem[{{Martins}, {Schaerer} \& {Hillier}(2005){Martins}, {Schaerer}, \&
  {Hillier}}]{Martins_etA_2005}
{Martins} F., {Schaerer} D., {Hillier} D.~J., 2005, \aap, 436, 1049

\bibitem[{{Mehner} {et~al}\mbox{.}(2010){Mehner}, {Davidson}, {Ferland}, \&
  {Humphreys}}]{Mehner_etA_2010}
{Mehner} A., {Davidson} K., {Ferland} G.~J., {Humphreys} R.~M., 2010, \apj,
  710, 729

\bibitem[{{Mehner} {et~al}\mbox{.}(2012){Mehner}, {Davidson}, {Humphreys},
  {Ishibashi}, {Martin}, {Ruiz}, \& {Walter}}]{Mehner_etA_2012}
{Mehner} A., {Davidson} K., {Humphreys} R.~M., {Ishibashi} K., {Martin} J.~C.,
  {Ruiz} M.~T., {Walter} F.~M., 2012, \apj, 751, 73

\bibitem[{{Mehner} {et~al}\mbox{.}(2011){Mehner}, {Davidson}, {Martin},
  {Humphreys}, {Ishibashi}, \& {Ferland}}]{Mehner_etA_2011}
{Mehner} A., {Davidson} K., {Martin} J.~C., {Humphreys} R.~M., {Ishibashi} K.,
  {Ferland} G.~J., 2011, \apj, 740, 80

\bibitem[{{Mellema} {et~al}\mbox{.}(2006){Mellema}, {Iliev}, {Alvarez}, \&
  {Shapiro}}]{Mellema_etA_2006}
{Mellema} G., {Iliev} I.~T., {Alvarez} M.~A., {Shapiro} P.~R., 2006, \na, 11,
  374

\bibitem[{{Monaghan}(1992)}]{Monaghan_1992}
{Monaghan} J.~J., 1992, \araa, 30, 543

\bibitem[{{Nielsen} {et~al}\mbox{.}(2007){Nielsen}, {Corcoran}, {Gull},
  {Hillier}, {Hamaguchi}, {Ivarsson}, \& {Lindler}}]{Nielsen_etA_2007}
{Nielsen} K.~E., {Corcoran} M.~F., {Gull} T.~R., {Hillier} D.~J., {Hamaguchi}
  K., {Ivarsson} S., {Lindler} D.~J., 2007, \apj, 660, 669

\bibitem[{{Okazaki} {et~al}\mbox{.}(2008){Okazaki}, {Owocki}, {Russell}, \&
  {Corcoran}}]{Okazaki_etA_2008}
{Okazaki} A.~T., {Owocki} S.~P., {Russell} C.~M.~P., {Corcoran} M.~F., 2008,
  \mnras, 388, L39

\bibitem[{{Owocki}(2007)}]{Owocki_2007}
{Owocki} S., 2007, in Astronomical Society of the Pacific Conference Series,
  Vol. 367, Massive Stars in Interactive Binaries, {St.-Louis} N., {Moffat}
  A.~F.~J., eds., p. 233

\bibitem[{{Paardekooper}(2010)}]{Paardekooper_2010}
{Paardekooper} J.-P., 2010, PhD thesis, Ph.~D.~thesis, University of Leiden
  (2010)

\bibitem[{{Paardekooper}, {Kruip} \& {Icke}(2010){Paardekooper}, {Kruip}, \&
  {Icke}}]{Paardekooper_etA_2010}
{Paardekooper} J.-P., {Kruip} C.~J.~H., {Icke} V., 2010, \aap, 515, A79

\bibitem[{{Paardekooper} {et~al}\mbox{.}(2011){Paardekooper}, {Pelupessy},
  {Altay}, \& {Kruip}}]{Paardekooper_etA_2011}
{Paardekooper} J.-P., {Pelupessy} F.~I., {Altay} G., {Kruip} C.~J.~H., 2011,
  \aap, 530, A87

\bibitem[{{Parkin} {et~al}\mbox{.}(2011){Parkin}, {Pittard}, {Corcoran}, \&
  {Hamaguchi}}]{Parkin_etA_2011}
{Parkin} E.~R., {Pittard} J.~M., {Corcoran} M.~F., {Hamaguchi} K., 2011, \apj,
  726, 105

\bibitem[{{Parkin} {et~al}\mbox{.}(2009){Parkin}, {Pittard}, {Corcoran},
  {Hamaguchi}, \& {Stevens}}]{Parkin_etA_2009}
{Parkin} E.~R., {Pittard} J.~M., {Corcoran} M.~F., {Hamaguchi} K., {Stevens}
  I.~R., 2009, \mnras, 394, 1758

\bibitem[{{Parkin} \& {Sim}(2013)}]{Parkin_Sim_2013}
{Parkin} E.~R., {Sim} S.~A., 2013, \apj, 767, 114

\bibitem[{{Pawlik} \& {Schaye}(2008)}]{Pawlik_Schaye_2008}
{Pawlik} A.~H., {Schaye} J., 2008, \mnras, 389, 651

\bibitem[{{Pittard}(2009)}]{Pittard_2009}
{Pittard} J.~M., 2009, \mnras, 396, 1743

\bibitem[{{Pittard} \& {Corcoran}(2002)}]{Pittard_Corcoran_2002}
{Pittard} J.~M., {Corcoran} M.~F., 2002, \aap, 383, 636

\bibitem[{{Price}(2007)}]{Price_2007}
{Price} D.~J., 2007, \pasa, 24, 159

\bibitem[{{Price}(2008)}]{Price_2008}
{Price} D.~J., 2008, Journal of Computational Physics, 227, 10040

\bibitem[{{Ritzerveld} \& {Icke}(2006)}]{Ritzerveld_Icke_2006}
{Ritzerveld} J., {Icke} V., 2006, \pre, 74, 026704

\bibitem[{{Ritzerveld}(2007)}]{Ritzerveld_2007}
{Ritzerveld} N.~G.~H., 2007, PhD thesis, Leiden Observatory, Leiden University,
  P.O.~Box 9513, 2300 RA Leiden, The Netherlands

\bibitem[{Russell(2013)}]{Russell_2013}
Russell C. M.~P., 2013, PhD thesis, University of Delaware, Newark, DE, USA

\bibitem[{{Smith}(2005)}]{Smith_2005}
{Smith} N., 2005, \mnras, 357, 1330

\bibitem[{{Smith}(2006)}]{Smith_2006}
{Smith} N., 2006, \apj, 644, 1151

\bibitem[{{Smith} {et~al}\mbox{.}(2003){Smith}, {Gehrz}, {Hinz}, {Hoffmann},
  {Hora}, {Mamajek}, \& {Meyer}}]{Smith_etA_2003a}
{Smith} N., {Gehrz} R.~D., {Hinz} P.~M., {Hoffmann} W.~F., {Hora} J.~L.,
  {Mamajek} E.~E., {Meyer} M.~R., 2003, \aj, 125, 1458

\bibitem[{{Steffen} {et~al}\mbox{.}(2014){Steffen}, {Teodoro}, {Madura},
  {Groh}, {Gull}, {Mehner}, {Corcoran}, {Damineli}, \&
  {Hamaguchi}}]{Steffen_etA_2014}
{Steffen} W. {et~al.}, 2014, \mnras, 442, 3316

\bibitem[{{Steiner} \& {Damineli}(2004)}]{Steiner_Daminelli_2004}
{Steiner} J.~E., {Damineli} A., 2004, \apjl, 612, L133

\bibitem[{{Stevens}, {Blondin} \& {Pollock}(1992){Stevens}, {Blondin}, \&
  {Pollock}}]{Stevens_etA_1992}
{Stevens} I.~R., {Blondin} J.~M., {Pollock} A.~M.~T., 1992, \apj, 386, 265

\bibitem[{{Stevens} \& {Pollock}(1994)}]{Stevens_Pollock_1994}
{Stevens} I.~R., {Pollock} A.~M.~T., 1994, \mnras, 269, 226

\bibitem[{{Teodoro} {et~al}\mbox{.}(2012){Teodoro}, {Damineli}, {Arias}, {de
  Ara{\'u}jo}, {Barb{\'a}}, {Corcoran}, {Borges Fernandes},
  {Fern{\'a}ndez-Laj{\'u}s}, {Fraga}, {Gamen}, {Gonz{\'a}lez}, {Groh},
  {Marshall}, {McGregor}, {Morrell}, {Nicholls}, {Parkin}, {Pereira},
  {Phillips}, {Solivella}, {Steiner}, {Stritzinger}, {Thompson}, {Torres},
  {Torres}, \& {Zevallos Herencia}}]{Teodoro_etA_2012}
{Teodoro} M. {et~al.}, 2012, \apj, 746, 73

\bibitem[{{Teodoro} {et~al}\mbox{.}(2008){Teodoro}, {Damineli}, {Sharp},
  {Groh}, \& {Barbosa}}]{Teodoro_etA_2008}
{Teodoro} M., {Damineli} A., {Sharp} R.~G., {Groh} J.~H., {Barbosa} C.~L.,
  2008, \mnras, 387, 564

\bibitem[{{Teodoro} {et~al}\mbox{.}(2013){Teodoro}, {Madura}, {Gull},
  {Corcoran}, \& {Hamaguchi}}]{Teodoro_etA_2013}
{Teodoro} M., {Madura} T.~I., {Gull} T.~R., {Corcoran} M.~F., {Hamaguchi} K.,
  2013, \apjl, 773, L16

\bibitem[{{Townsend}(2009)}]{Townsend_2009}
{Townsend} R.~H.~D., 2009, \apjs, 181, 391

\bibitem[{{Verner}, {Bruhweiler} \& {Gull}(2005){Verner}, {Bruhweiler}, \&
  {Gull}}]{Verner_etA_2005}
{Verner} E., {Bruhweiler} F., {Gull} T., 2005, \apj, 624, 973

\bibitem[{{Weigelt} \& {Ebersberger}(1986)}]{Weigelt_Ebersberger_1986}
{Weigelt} G., {Ebersberger} J., 1986, \aap, 163, L5

\end{thebibliography}

\label{lastpage}

\end{document}